%% file: Arxiv_Ver.tex
\documentclass{article} % For LaTeX2e
\usepackage{iclr2027_conference,times}

\input{math_commands.tex}

\usepackage{hyperref}
\usepackage{url}
\usepackage{graphicx}
\usepackage{amsmath}
\usepackage{amssymb}

\usepackage{url}

\usepackage{booktabs}
\usepackage{microtype}
\usepackage{graphicx}
\usepackage{booktabs} % for professional tables
\usepackage{makecell}
\usepackage{amsmath}

\usepackage{verbatim}
\usepackage{epstopdf}
\usepackage{multirow}
\usepackage{amsfonts}
\usepackage{soul}
\usepackage{graphicx}
\usepackage{subcaption} % 不要和 subfigure 宏包同时用
\usepackage{newfloat}
\usepackage{listings}
\usepackage[table]{xcolor}
\usepackage{wrapfig}
\usepackage[most]{tcolorbox}

\newtcolorbox[
    auto counter,
    number within=section
]{promptbox}[2][]{
    colback=gray!3,
    colframe=black!70,
    boxrule=0.8pt,
    arc=2mm,
    left=6pt,
    right=6pt,
    top=6pt,
    bottom=6pt,
    fonttitle=\bfseries,
    coltitle=black,
    colbacktitle=gray!10,
    title={Prompt~\thetcbcounter: #2},
    #1
}
\title{%Improving Role Adherence in Multi-Agent Collaboration via Dynamic Activation Steering
Mind Your Own Role: Reducing Role Overstepping for Effective and Efficient Multi-Agent Collaboration via Dynamic Role Activation Steering
}
\author{
\textbf{Xingke Yang$^{1}$,
Sicong Li$^{1}$,
Lingfeng Yao$^{1}$,
Rui Chen$^{2}$,
Gen Li$^{1}$,
Hao Wang$^{3}$
}
\\
\textbf{
Xin Fu$^{1}$,
Miao Pan$^{1}$
}
\\[0.5em]
$^{1}$University of Houston, USA \quad
$^{2}$Independent Researcher, USA \quad
\\
$^{3}$Stevens Institute of Technology, USA
}
\iclrfinalcopy % Uncomment for camera-ready version, but NOT for submission.
\begin{document}

\maketitle
\lhead{Preprint}

\begin{abstract}
%resulting in role overstepping, 
Multi-Agent Systems (MAS) have shown strong capability in solving complex tasks by assigning different agents specialized roles and corresponding subtasks. However, agents may overstep their assigned responsibilities and perform subtasks belonging to other roles, which can disrupt workflow, reduce system efficiency, and degrade task performance. To address this issue, we investigate role-related patterns in the model's internal activation space and uncover two key observations: (i) role instructions induce consistent directional shifts in activation space, and (ii) role-adherent and role-overstepping trajectories exhibit distinct activation patterns. Based on these observations, we propose Dynamic Role Activation Steering (DRAS), a lightweight and effective inference-time activation intervention approach to reduce role overstepping in multi-agent collaboration. Specifically, DRAS constructs a role steering vector and a lightweight role-overstepping detector for each role. During LLM generation, the detector estimates the probability of role overstepping from the current activation state, and DRAS dynamically adjusts the steering strength accordingly to guide the agent toward its assigned responsibilities. Experimental results show that DRAS effectively reduces the number of role-overstepping trajectories by up to 86.6\% compared with Vanilla MAS, thereby improving task quality score by 10.0\%--11.1\%, achieving a 1.50$\times$--1.61$\times$ task-completion speedup, and reducing token consumption by 26.8\%--38.7\%.

\end{abstract}

% Experimental results show that, by appropriately splitting workload between PEFT and RAG, PUR consistently achieves the best generation accuracy/quality for various users across multiple tasks, while reducing inference latency by 14.4\%--42.5\% compared with the best performed SOTA method.

\section{Introduction}

\textcolor{black}{In recent years, LLM-driven multi-agent systems have made significant progress in solving complex tasks~\citep{ye2025mas,hu2025automated,chen2024agentverse}, particularly for highly collaborative applications like software development. By equipping agents with distinct role profiles and defining interaction protocols, these systems decompose the long-horizon task into coordinated subtasks, allowing different agents to handle different subtasks while jointly completing the overall task. Most well-known multi-agent systems (e.g., ChatDev~\citep{qian2024chatdev} and MetaGPT~\citep{hong2024metagpt} and AutoGen~\citep{wu2023autogen}) employ such structured collaboration to improve reliability and end-to-end task performance.}

\begin{figure*}[t]
    \centering

    \begin{subfigure}[t]{0.6\linewidth}
        \centering
        \includegraphics[width=\linewidth]{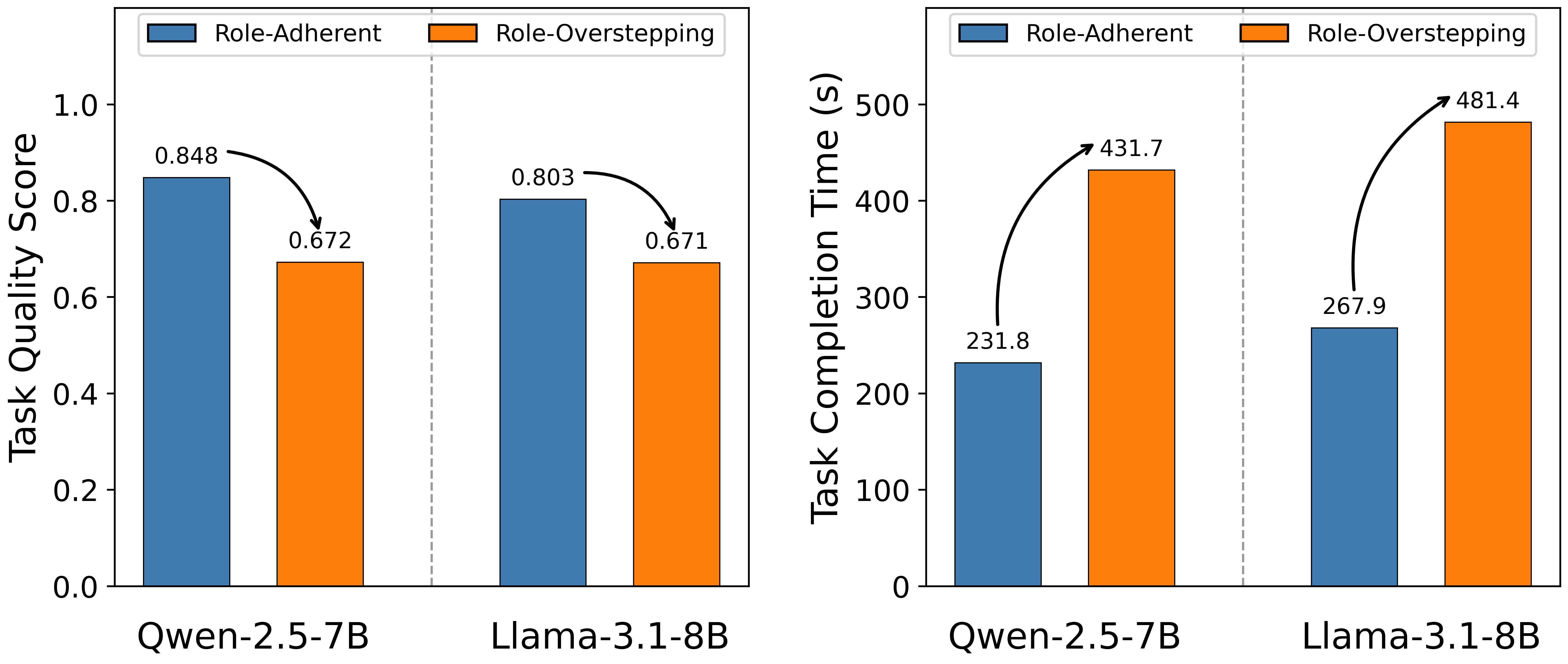}
        \caption{Task quality score and task completion time.}
        \label{fig:motivation_latency}
    \end{subfigure}
    \hfill
    \begin{subfigure}[t]{0.36\linewidth}
        \centering
        \includegraphics[width=\linewidth]{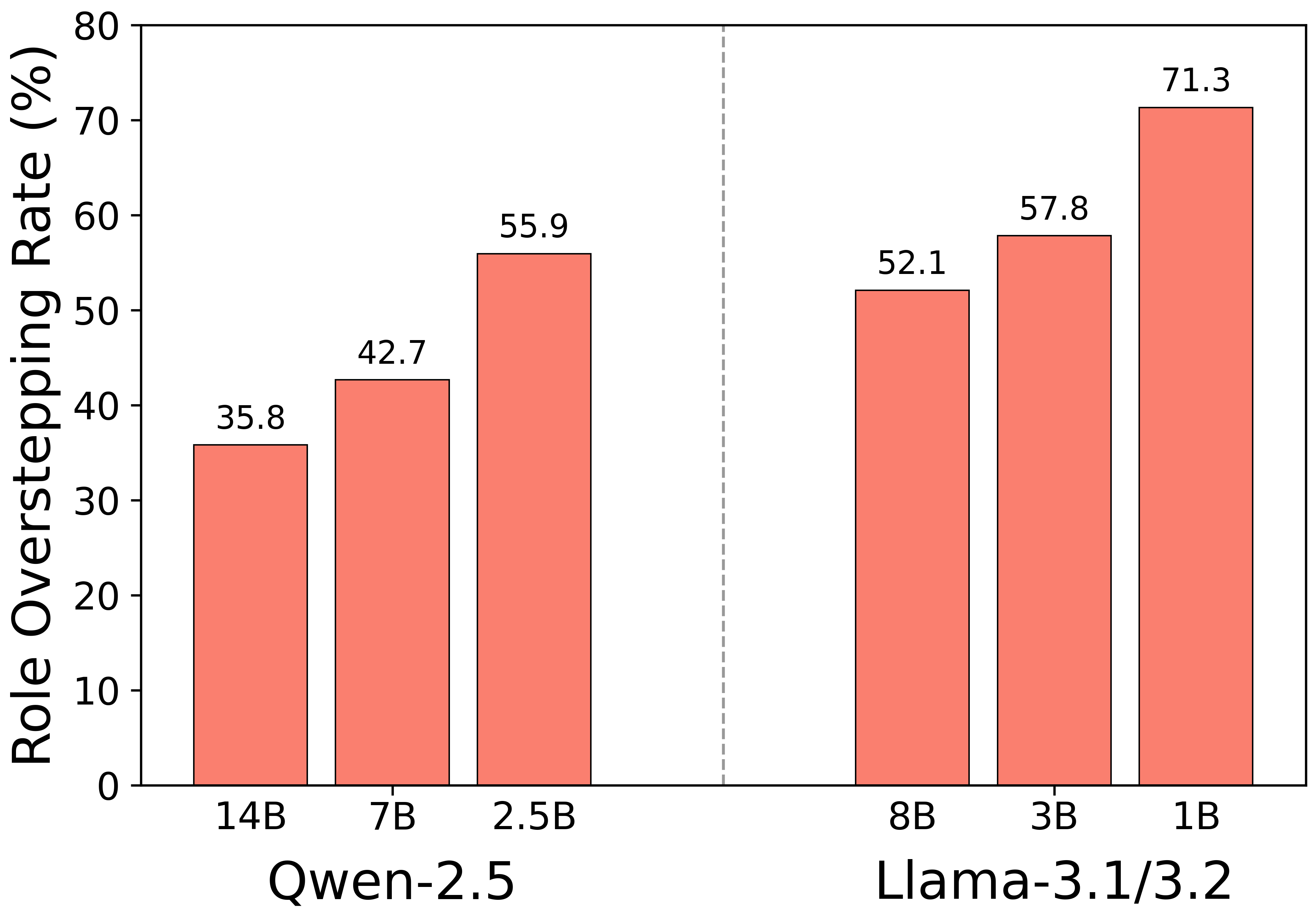}
        \caption{Role overstepping rates.}
        \label{fig:motivation_rate}
    \end{subfigure}

    \caption{\textcolor{black}{ Prevalence and impact of role overstepping in ChatDev. (a) Comparison of task performance and task completion time between role-adherent and role-overstepping tasks. (b) Role overstepping rates across model sizes.}}
    \label{fig:motivation_overview}
\end{figure*}

\textcolor{black}{However, recent studies~\citep{kim2026teambench,cemri2026multi,rath2026agent} have shown that in practice it does not always hold that each agent always follows its assigned role. Agents may exceed their predefined responsibilities and perform subtasks that belong to other roles, resulting in \textbf{role overstepping}~\citep{zhou2026improving}. For example, an agent responsible for code review may go beyond providing review comments and directly modify the code. Unlike an isolated generation error, the content generated by role overstepping may propagate along the collaboration chain, and cause redundant execution, responsibility conflicts, and contamination of the shared context. Consequently, role overstepping may degrade the overall task performance and increase the task completion time~\citep{kim2026teambench,rath2026agent}. As shown in Fig.~\ref{fig:motivation_latency}, \textcolor{black}{for} ChatDev with Qwen-2.5-7B ~\citep{Yang2024Qwen25TR}, tasks with role overstepping exhibit a 20.75\% decrease in task quality scores and an 86.24\% increase in task completion time compared with role-adherent tasks. Moreover, our empirical results in Fig.~\ref{fig:motivation_rate} show that role overstepping is prevalent across various model sizes and becomes more pronounced as the model size decreases.} \textcolor{black}{For example, even for Qwen-2.5-14B, the role-overstepping rate reaches 35.8\%, and further increases to 55.9\% for the 2.5B model. With ever growing interest of deploying multi-agent systems on resource-constrained mobile devices that typically run medium or small sized models~\citep{fu2025camphor, yi2026ecoagent, zhou2026mobile}, such a high role-overstepping rate and its consequent task performance degradation and low system efficiency become an essential problem.}

\textcolor{black}{To address this issue, ~\citet{wang2026detecting} examine agent outputs after generation, using rule-based detectors or an additional LLM-as-a-Judge to determine whether a response violates the assigned role, and triggers regeneration when overstepping is detected. However, such post-hoc methods can only identify overstepping after a response has already been generated, wasting the computation spent on the overstepping response. The detection and regeneration process also introduces substantial computational cost and task completion time. Alternatively, \citet{zhou2026improving} leverage role-adherent trajectories to perform LoRA-based LLM fine-tuning~\citep{hu2021lora} to improve role adherence. However, fine-tuning the LLM for each role incurs substantial training cost and hardware requirements, limiting its practicality and deployment flexibility. Therefore, how to promptly detect role overstepping and efficiently steer agent behavior remains an important challenge.}

 \textcolor{black}{Activation engineering has recently emerged as an effective approach to steering LLM behavior at inference time~\citep{rimsky2024steering,wehner2025taxonomy,valentino2026mitigating}. Its key idea is to identify directions in the model's activation space associated with specific behavioral attributes, such as honesty~\citep{qiu2024spectral}, truthfulness~\citep{li2023inference}, and humor~\citep{von2024language}, and to derive steering vectors from these directions. Instead of modifying model parameters, these methods directly intervene in internal activations using the extracted steering vectors, enabling cost-efficient behavioral control with adjustable steering strength. Inspired by this, we propose to study \textbf{if it is possible to analyze and intervene in the model's activation space during generation, so that we can promptly detect role-overstepping and efficiently fix it via activation steering.} To answer this question, we conduct exploratory experiments and reveal two key observations.  \textcolor{black}{(i) Role instructions induce consistent directional shifts in activation space, and steering along these directions can be viewed as reinforcing the behavioral constraints specified by the role instructions.} (ii) Role-adherent and role-overstepping behaviors exhibit distinct activation patterns. These observations indicate that the activation space contains both the directional information needed for role steering and the state information needed for role-overstepping detection.}

% \textcolor{purple}{Activation engineering has recently emerged as an effective approach for controlling LLM behavior at inference time~\citep{rimsky2024steering,wehner2025taxonomy,valentino2026mitigating}. Instead of modifying model parameters, these methods directly intervene in internal activations and have been applied to behavioral attributes such as harmlessness, truthfulness, and humor. Motivated by this, we investigate a key question: \textbf{Can role information also be encoded in the internal activation space of LLMs in a form that supports both role-behavior detection and steering?} We conduct exploratory experiments and reveal two key observations. First, role instructions induce stable directional shifts in activation space, which can be extracted as role-specific steering directions. Second, role-adherent and role-overstepping behaviors exhibit clear separability in activation space. These observations suggest that internal activations provide both a signal for detecting emerging role overstepping and a direction for steering the model toward role-adherent behavior.}

 \textcolor{black}{Based on those observations above, in this paper, we propose Dynamic Role Activation Steering (DRAS), an inference-time activation intervention approach for reducing role overstepping in multi-agent collaboration. For each role, we (i) compare the internal activations produced by the same subtask context with and without the corresponding role instruction and derive a role steering vector from their activation differences, and (ii) collect activations from role-adherent and role-overstepping trajectories to train a lightweight role-overstepping detector. During LLM generation, the detector estimates the probability of role overstepping from the current activation representation, which is then used to dynamically adjust the steering strength and promote role adherence. }\textcolor{black}{Extensive experiments demonstrate that DRAS reduces the number of role-overstepping trajectories by up to 86.6\% compared with Vanilla MAS and consistently outperforms other methods. By reducing role overstepping, DRAS further improves task quality score by 10.0\%--11.1\%, achieves a 1.50$\times$--1.61$\times$ task-completion speedup, and reduces token consumption by 26.8\%--38.7\%. }

\section{Preliminaries}

\subsection{Role Overstepping in Multi-Agent Collaboration}

\textcolor{black}{Unlike end-to-end evaluation of overall task performance in a multi-agent system, role overstepping concerns whether each agent properly executes its assigned role within the internal collaborative workflow. In role-based multi-agent systems, agent responsibilities are usually defined by natural-language role instructions in LLM prompts. Formally, consider a multi-agent system with a role set $\mathcal{R}=\{r_1,\ldots,r_K\}$. Each role $r_i$ is associated with a role instruction $I_{r_i}$, which specifies its responsibilities and behavioral constraints. Given a subtask context $x_i$, the agent with role $r_i$ generates a behavior trajectory $\tau_i=\mathrm{LLM}(x_i,I_{r_i})$.}

\textcolor{black}{\textbf{Definition:} Role $r_i$ exhibits role overstepping if $\tau_i$ contains behaviors that fall within the responsibilities described in another role $r_j$'s role instruction $I_{r_j}$.}

\textcolor{black}{For trajectory evaluation, existing studies~\citep{wang2026detecting,kim2026teambench} commonly adopt LLM-as-a-Judge to determine whether each trajectory exhibits role overstepping according to the predefined role division. In addition, ~\cite{zhou2026improving} introduce a continuous metric based on semantic similarity in the embedding space. By comparing a trajectory with different role descriptions, the metric quantifies its alignment with the assigned role and its degree of shift toward other roles.}

\subsection{Activation Engineering}

\textcolor{black}{\textbf{Activation Steering:} Emerging research in activation engineering has uncovered that LLMs encode semantic concepts as linear subspaces within internal activations~\citep{liu2023context,zou2023representation,arditi2024refusal}. This geometric interpretation enables targeted behavioral steering through linear interventions during inference. ~\citet{rimsky2024steering} pioneered activation addition using steering vectors derived from contrastive activations for sentiment and topic control. Following the same principle, ~\citet{konen2024style} demonstrated that activation vectors can also steer linguistic style, while ~\cite{zhang2024truthx} further identified steerable activation directions associated with the truthfulness of model outputs. More recently, ~\cite{liu2026perfit, zhang2025personalized} extracted activation shifts associated with user personal information to guide personalized generation. In this work, we investigate whether role-specific behaviors exhibit identifiable directional representations in activation space that can be leveraged for role steering in multi-agent collaboration.}

\textcolor{black}{\textbf{Conditional Activation Steering:} Applying a steering vector indiscriminately may interfere with model behaviors that do not require correction. For example, strengthening refusal to harmful queries can also suppress appropriate responses to harmless queries~\citep{lee2025programming}. Therefore, a fine-grained strategy is to condition the steering intervention on the model's current behavioral state. ~\cite{lee2025programming} uses activation differences between harmful and harmless prompts to construct a conditioning vector that determines whether steering should be activated.  ~\cite{wang2025adaptive,zhang2024truthx} use labeled activations from truthful and untruthful trajectories to train a lightweight linear probe, which detects the truthfulness state of the current generation and adjusts steering strength accordingly.  These studies demonstrate that activation space not only contains directional representations for behavioral steering, but also encodes information that reflects the model's current behavioral state.}

\section{Uncovering Role-Related Patterns in Activation Space}

\textcolor{black}{Building on the insights of activation steering, we investigate whether role-related information exhibits structured and identifiable patterns in the model's internal activation space. Specifically, we extract model activations under contrastive prompts and use principal component analysis (PCA)\footnote{Principal component analysis (PCA) is a linear dimensionality reduction technique. It linearly projects the data onto a new coordinate system, where the axes (principal components) are selected to account for the most significant variance in the data.} to assess the separability of their internal representations. This analysis is useful for examining whether activation differences under contrastive conditions exhibit consistent and identifiable structures~\citep{park2023linear}}.

\begin{figure*}[t]
    \centering

    % ---------- (a) Left two panels ----------
    \begin{subfigure}[t]{0.49\linewidth}
        \centering
        \includegraphics[width=0.49\linewidth]{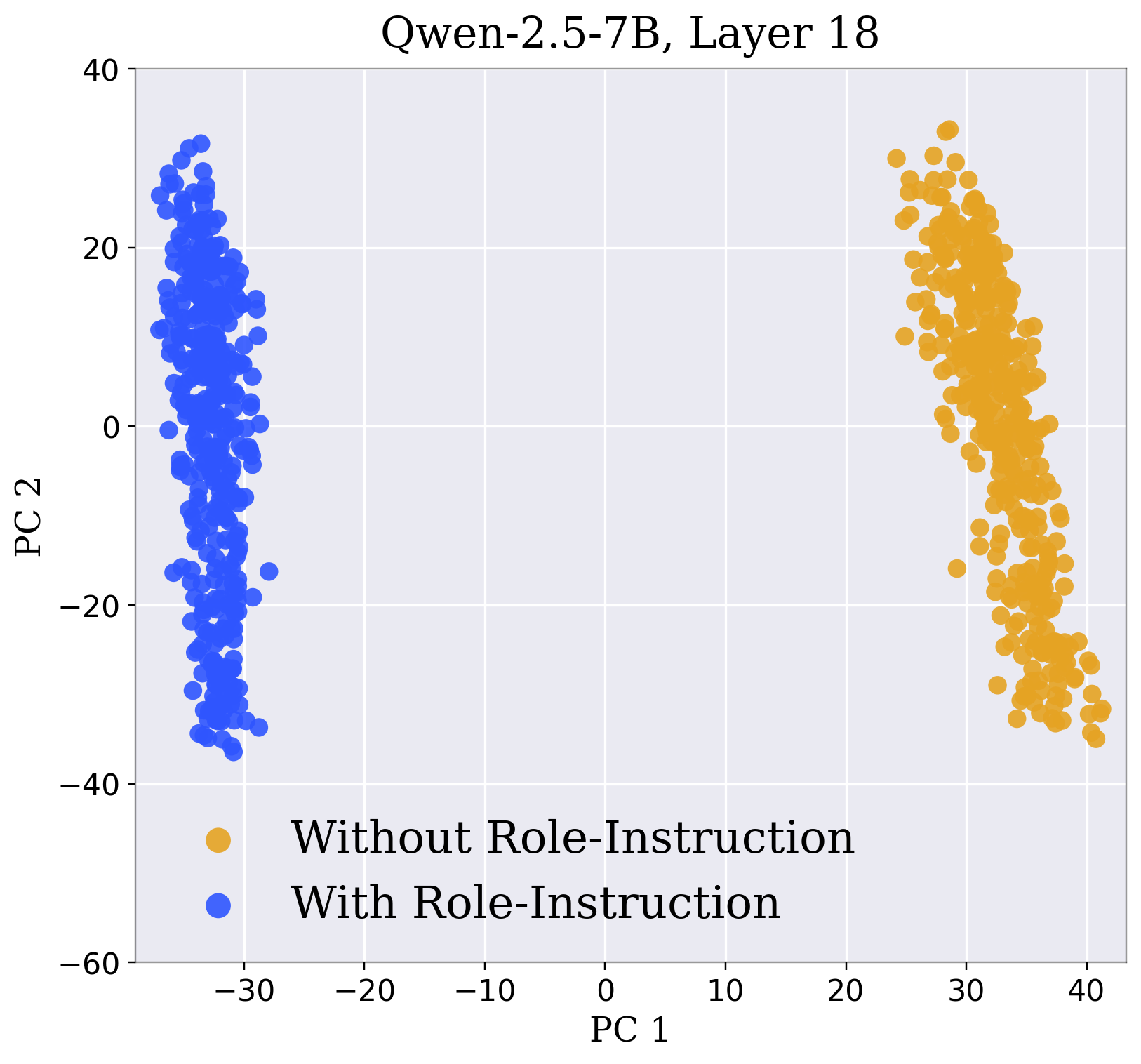}
        \hfill
        \includegraphics[width=0.49\linewidth]{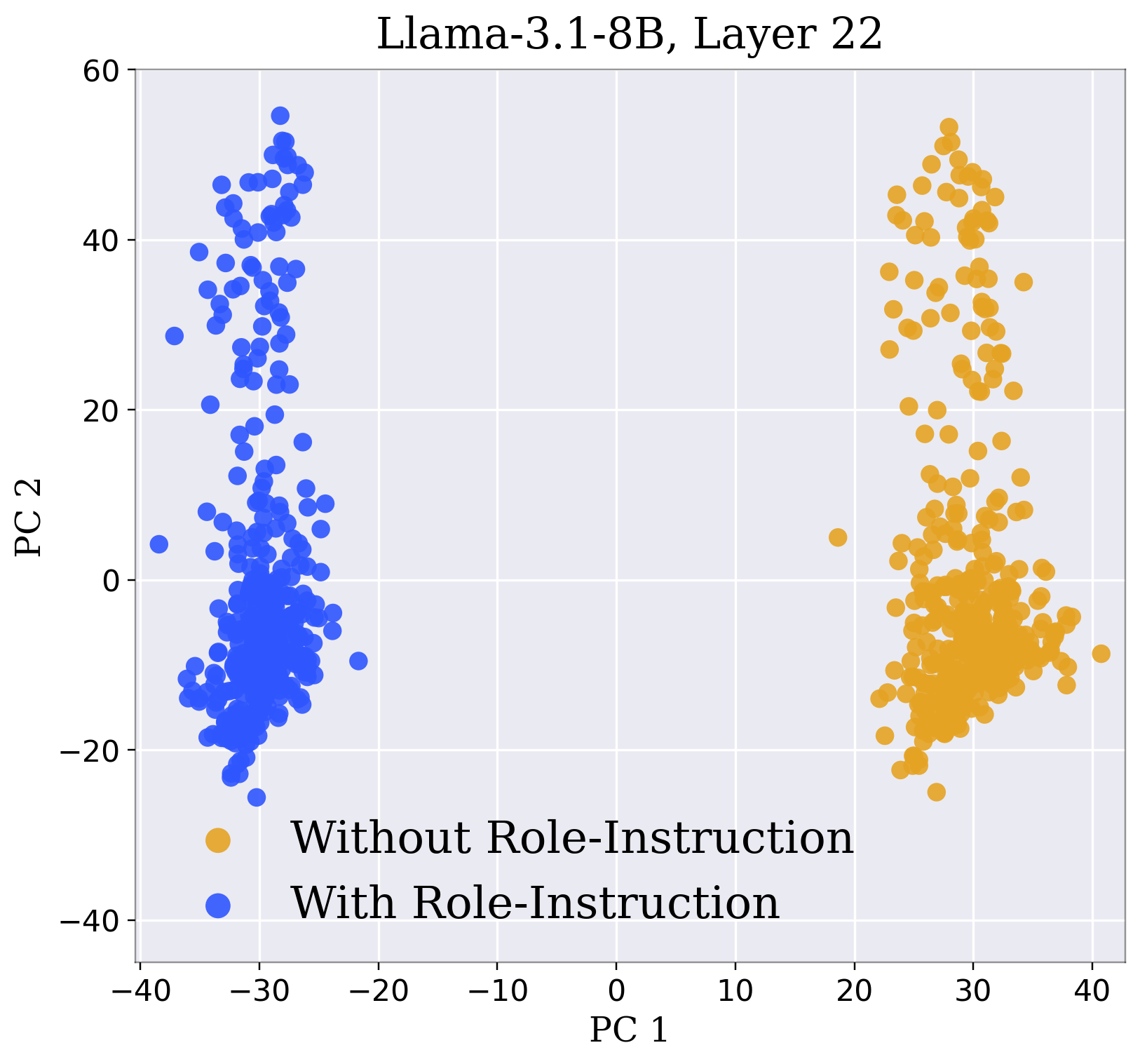}
        \caption{PCA projections of activations from prompts with and without role instructions.}
        \label{fig:PCA1}
    \end{subfigure}
    \hfill
    % ---------- (b) Right two panels ----------
    \begin{subfigure}[t]{0.49\linewidth}
        \centering
        \includegraphics[width=0.49\linewidth]{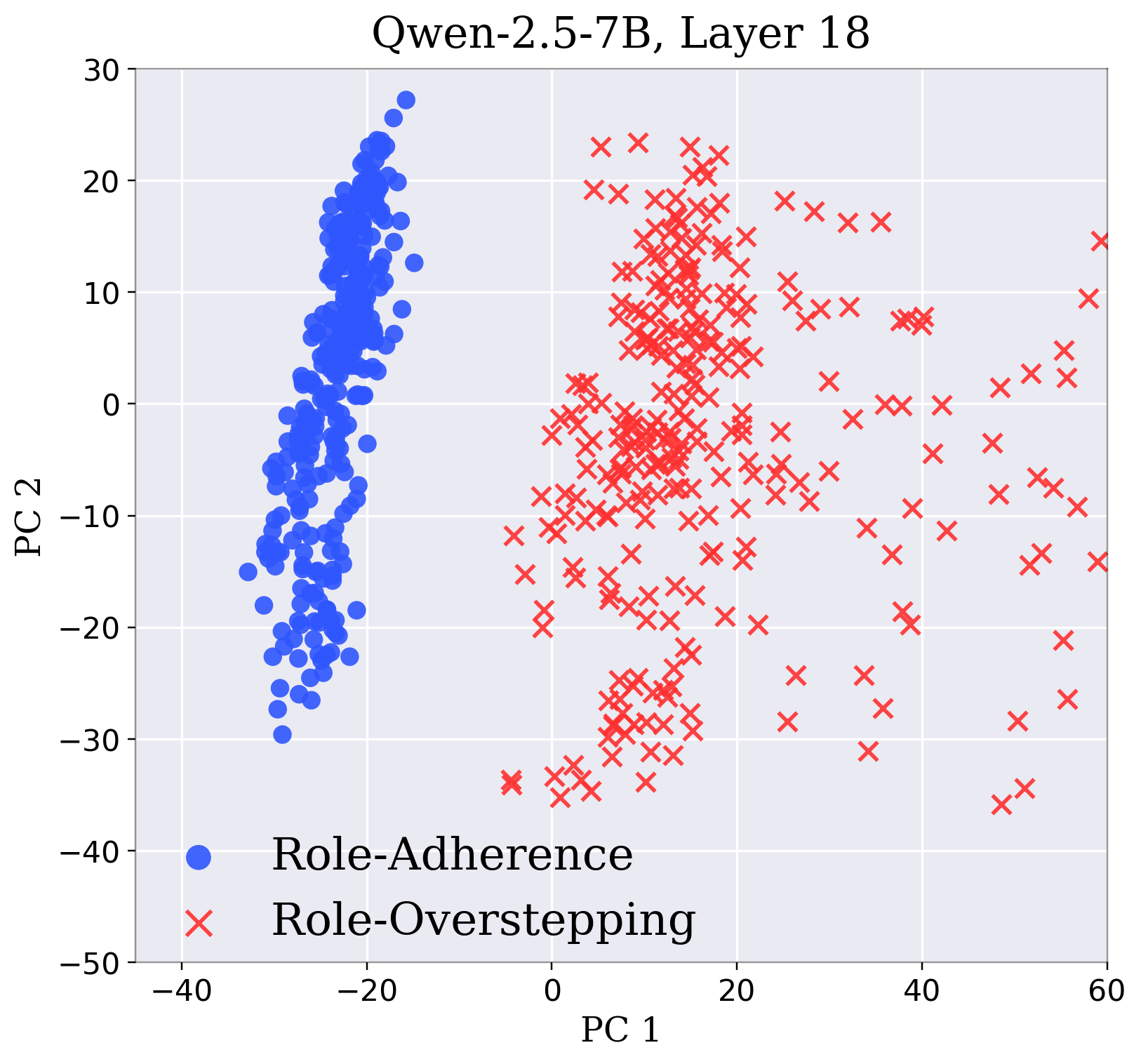}
        \hfill
        \includegraphics[width=0.49\linewidth]{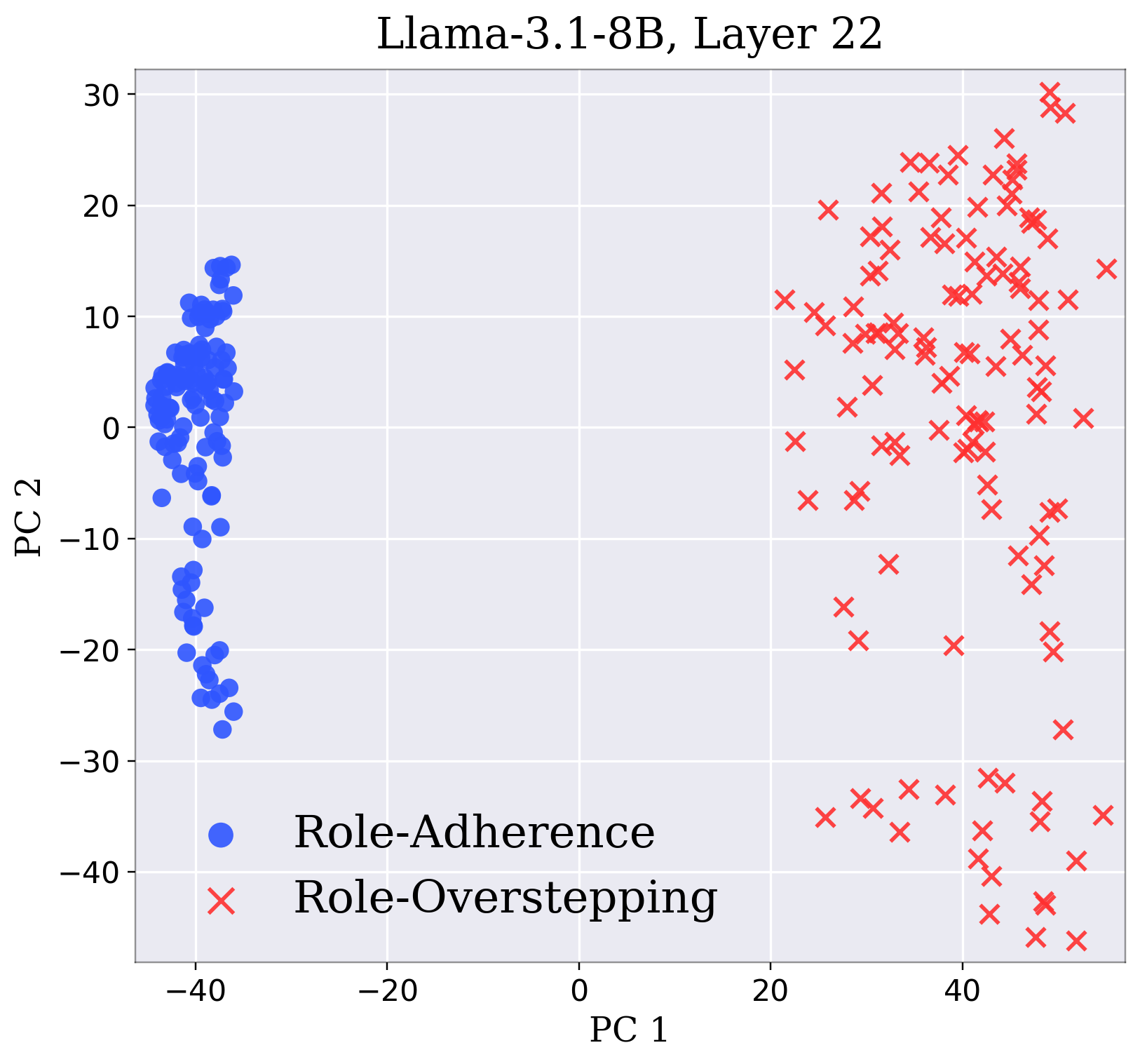}
        \caption{PCA projections of activations from role-adherent and role-overstepping trajectories.}
        \label{fig:PCA2}
    \end{subfigure}

    \caption{
        \textcolor{black}{PCA projections of internal activations in ChatDev for the CEO role, extracted at selected layers of Qwen2.5-7B and Llama-3.1-8B.}
    }
    \label{fig:PCA}
\end{figure*}

\textcolor{black}{\textbf{Observation 1: Role instructions induce consistent activation shifts.} We form contrastive prompt pairs using the same role subtask context with and without the corresponding role instruction. For each pair, we perform separate forward passes and extract the last-token residual-stream activations from the same model layer. As shown in Fig.~\ref{fig:PCA1}, adding the role instruction produces a clear and consistent shift in the activation distribution at the selected layers. It indicates that role instructions induce systematic activation shifts that may provide an extractable steering direction for role-specific behaviors.}

% \textcolor{purple}{\textbf{Observation 2: Role-adherent and Role-overstepping behaviors are separable.} We construct two contrastive trajectory groups consisting of role-adherent and role-overstepping trajectories generated for the same agent role, and extract their last-token residual-stream activations from the same model layer. As shown in Fig. 2(b), the two behavioral states exhibit clearly distinct activation distributions around the middle layers. This suggests that an agent's role-adherence state can be detected from its activation representations during generation.}

\textcolor{black}{\textbf{Observation 2: Role-adherent and role-overstepping behaviors exhibit distinct activation patterns.} We collect role-adherent and role-overstepping trajectories generated for the same agent role and extract their last-token residual-stream activations from the same model layer. As shown in Fig.~\ref{fig:PCA2}, the two behavioral states exhibit clearly distinct activation distributions in the PCA projections at the selected layers. It means that internal activations contain discriminative information about whether an agent is adhering to or overstepping its assigned role.}

 % We collect role-adherent and role-overstepping trajectories generated for the same agent role and extract their last-token residual-stream activations from the same model layer. As shown in Fig. 2(b), the two behavioral states exhibit clearly distinct activation distributions in the PCA projections at the selected layers. This suggests that role-adherent and role-overstepping behaviors can be distinguished from their internal activation representations.

\section{Method}

\textcolor{black}{Based on the observations above, we propose Dynamic Role Activation Steering (DRAS), an inference-time activation intervention approach for mitigating role overstepping in multi-agent collaboration. For each role, alongside the original role instruction, we introduce a dynamic steering module consisting of a role steering vector and a role-overstepping detector. During generation, the module estimates the probability of role overstepping from activations at a selected LLM layer and dynamically adjusts the steering strength to effectively reduce role overstepping.}

\subsection{Role Steering Vector Construction}\label{sec:role_steering_vector}

\textcolor{black}{Following the contrastive activation paradigm, we construct a role steering vector from activation differences induced by the corresponding role instruction. For each role $r$, we consider $N$ subtask contexts $\{x_n\}_{n=1}^{N}$. For each $x_n$, we construct a matched input pair consisting of the subtask context alone, i.e., $x_n$, and the same context augmented with the role instruction, i.e., $[x_n; I_r]$. Let $h^{\emptyset}_{n,l} \in \mathbb{R}^{d}$ denote the last-token residual-stream activation at the output of layer $l$ for $x_n$, and $h^{r}_{n,l} \in \mathbb{R}^{d}$ denote that for $[x_n; I_r]$, respectively. The role steering vector $\mathbf{v}^{l}_{r}$ for role \(r\) at layer \(l\) is defined as}
\begin{equation}
\mathbf{v}^{l}_{r}
=
\frac{1}{N}
\sum_{n=1}^{N}
\left(
h^{r}_{n,l}
-
h^{\emptyset}_{n,l}
\right).
\label{eq:role_vector}
\end{equation}
\textcolor{black}{Averaging across different subtask contexts reduces task-specific variation and captures the consistent activation shift introduced by the role instruction. In this work, we extract activations and apply steering at a single layer $l$, which is selected based on probe performance on the validation set.}

% \begin{equation}
% \mathbf{v}^{l}_{r}
% =
% \frac{1}{N}
% \sum_{n=1}^{N}
% \left(
% h^{+}_{n,l}
% -
% h_{n,l} \right).
% \label{eq:role_vector}
% \end{equation}

\subsection{Role-Overstepping Detector Training}\label{sec:Detector_train}

\textcolor{black}{To enable fine-grained role steering and avoid unnecessary activation perturbations on role-adherent generations, we control the steering strength according to the estimated probability of role overstepping. Linear probes are commonly used to understand information encoded in internal representations~\citep{belinkov2022probing,wang2025adaptive}. Based on Observation 2, we employ a lightweight linear probe to differentiate the activation representations of role-adherent and role-overstepping behaviors.}

\textcolor{black}{For each role $r$, we collect $N_{\mathrm{adh}}$ role-adherent trajectories $\{\tau_n^{\mathrm{adh}}\}_{n=1}^{N_{\mathrm{adh}}}$ and $N_{\mathrm{over}}$ role-overstepping trajectories $\{\tau_m^{\mathrm{over}}\}_{m=1}^{N_{\mathrm{over}}}$. 
Then we construct two types of inputs by concatenating
the subtask context, role instruction $I_r$, and trajectory, $[x_n; I_r; \tau_n^{\mathrm{adh}}]$ and $[x_m; I_r; \tau_m^{\mathrm{over}}]$, respectively. Let $h_{n,l}^{\mathrm{adh}} \in \mathbb{R}^{d}$ and $h_{m,l}^{\mathrm{over}} \in \mathbb{R}^{d}$ denote their last-token residual-stream activations at the output of layer $l$, respectively. We assign label $0$ to role-adherent activations and label $1$ to role-overstepping activations, and form the probing dataset as follows.}

\begin{equation}
\mathcal{D}_r^{\mathrm{probe}}
=
\left\{
\left(h_{n,l}^{\mathrm{adh}},0\right)
\right\}_{n=1}^{N_{\mathrm{adh}}}
\cup
\left\{
\left(h_{m,l}^{\mathrm{over}},1\right)
\right\}_{m=1}^{N_{\mathrm{over}}}.
\label{eq:probe_dataset}
\end{equation}

% \textcolor{purple}{For each role $r$, we collect $N$ role-adherent trajectories $\tau_n^{\mathrm{adh}}$ and $N$ role-overstepping trajectories $\tau_n^{\mathrm{over}}$. 
% Then we construct two types of inputs by concatenating
% the task context, role instruction $I_r$, and trajectory, $[x_n; I_r; \tau_n^{\mathrm{adh}}]$ and $[x_n; I_r; \tau_n^{\mathrm{over}}]$. Let $h_{n,l}^{\mathrm{adh}} \in \mathbb{R}^{d}$ and $h_{n,l}^{\mathrm{over}} \in \mathbb{R}^{d}$ denote their residual-stream activations at the last input token of layer $l$, respectively. We assign label \(0\) to role-adherent activations and label \(1\) to role-overstepping activations, forming the probing dataset:}

% \begin{equation}
% \left\{
% \left(h_{n,l}^{\mathrm{adh}}, 0\right),
% \left(h_{n,l}^{\mathrm{over}}, 1\right)
% \right\}_{n=1}^{N},
% \label{eq:probe_dataset}
% \end{equation}

% Using this dataset, we independently train the linear probe with a binary cross-entropy objective to distinguish role-adherent from role-overstepping activation states, requiring only minimal training overhead.

\textcolor{black}{Using this dataset, we independently train the linear probe with a binary cross-entropy objective to differentiate role-adherent activation representations from role-overstepping ones, which has very limited training overhead. The detector for role $r$ at layer $l$ is defined as}

\begin{equation}
D_r^{l}(\mathbf{h})
=
\sigma\bigl(
(\mathbf{w}_r^{l})^{\top}\mathbf{h}
+
b_r^{l}
\bigr),
\label{eq:role_detector}
\end{equation}

\textcolor{black}{where $\sigma(\cdot)$ is the sigmoid function, and $\mathbf{w}_r^{l} \in \mathbb{R}^{d}$ and $b_r^{l}$ are the trainable parameters of the detector. The output $D_r^{l}(\mathbf{h}) \in [0,1]$ represents the estimated probability of role overstepping for the current activation representation.}

\begin{figure*}[t]
    \centering
    \includegraphics[width=0.99\textwidth]{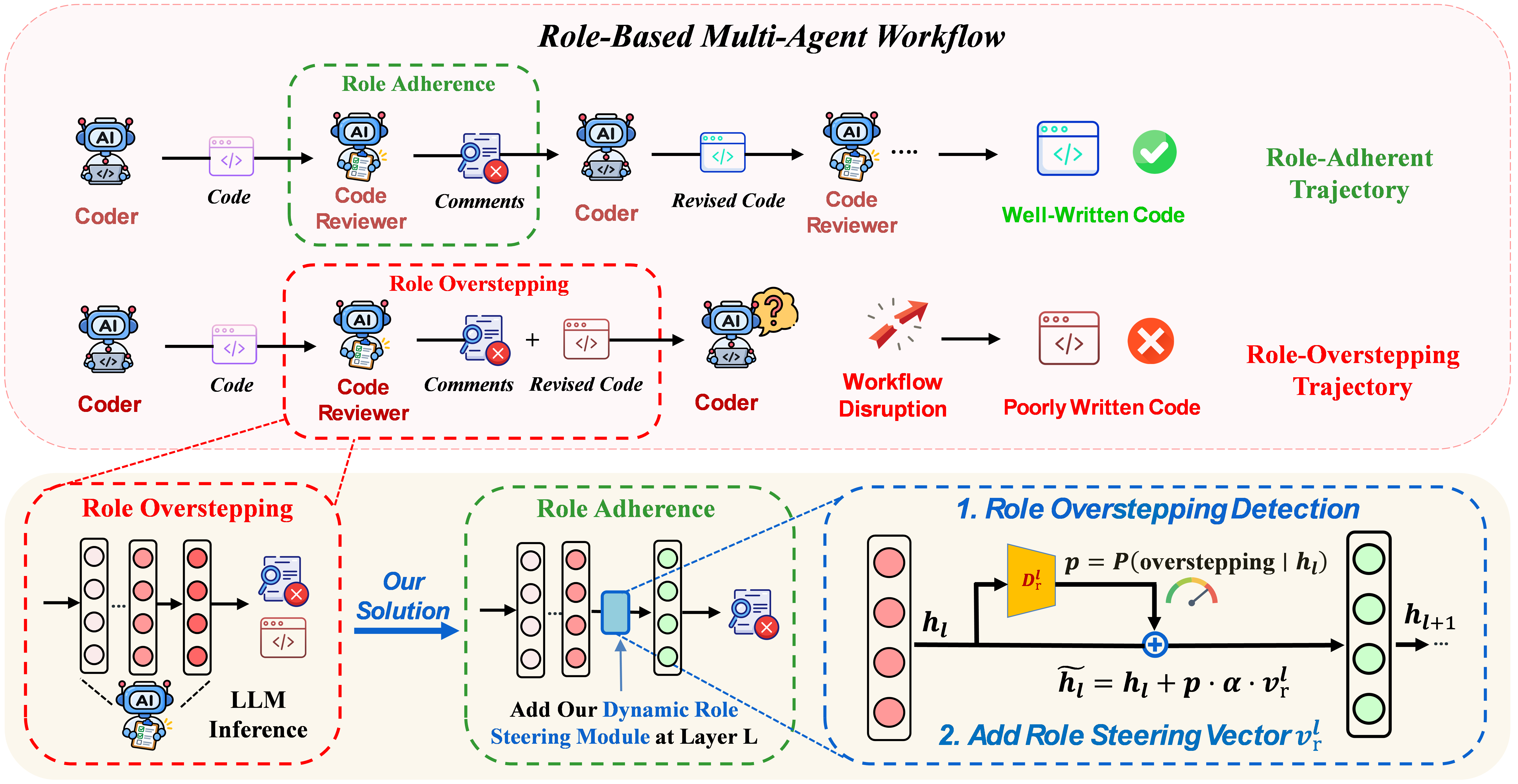}
    \caption{\textcolor{black}{Role overstepping in multi-agent collaboration and our proposed Dynamic Role Activation Steering (DRAS). During generation, DRAS first detects the role-overstepping probability from the current activation state and then dynamically adjusts the steering strength to guide the agent toward its assigned role.}}
    \label{fig:overview}
    % \vspace{-0.5em}
\end{figure*}

\subsection{Dynamic Steering During Generation}

\textcolor{black}{After constructing the role steering vector and role-overstepping detector, we perform dynamic activation steering during LLM generation. Following ~\cite{rimsky2024steering}, we apply the intervention at every decoding step. As illustrated in Fig.~\ref{fig:overview}, let $\mathbf{h}_l$ denote the residual-stream activation at the last token of the current sequence at the selected layer $l$. We first feed $\mathbf{h}_l$ into the role-overstepping detector to obtain an estimated probability of role overstepping $p=D_r^l(\mathbf{h}_l)\in[0,1]$. We then update the activation as}

% \textcolor{purple}{After constructing the role steering vector and role-overstepping detector, we perform dynamic activation steering during LLM generation. Following ~\cite{rimsky2024steering}, we apply the intervention at every decoding step after the initial prompt $[x; I_r]$ has been processed. As illustrated in Fig.~\ref{fig:overview}, at each decoding step, let $\mathbf{h}_l$ denote the residual-stream activation at the last token of the current sequence at the selected layer $l$. We first feed $\mathbf{h}_l$ into the role-overstepping detector to obtain an estimated probability of role overstepping $p=D_r^l(\mathbf{h}_l)\in[0,1]$. We then update the activation as}

\begin{equation}
\tilde{\mathbf{h}}_l
=
\mathbf{h}_l
+
\alpha p\,\mathbf{v}_r^l,
\end{equation}

\textcolor{black}{where $\alpha$ is a hyperparameter that controls the base steering strength, while $p$ dynamically scales the intervention according to the estimated probability of role overstepping. The modified activation $\tilde{\mathbf{h}}_l$ is then passed through the subsequent Transformer computation. In this way, our DRAS continuously monitors the agent's role behavior and dynamically applies the role activation steering, which keeps reducing role overstepping behaviors throughout the generation process.}

\section{Experimental Setup}\label{sec:Experimental Setup}

\textcolor{black}{\noindent{\bf Multi-Agent Systems (MAS) and LLM Backbones:} We evaluate DRAS on two representative role-based multi-agent systems, ChatDev~\citep{qian2024chatdev} and MetaGPT~\citep{hong2024metagpt}, both of which organize multiple LLM agents with predefined role responsibilities to collaboratively complete software-development tasks. For each system, we preserve its original workflow and role instructions. For the main experiments, we use two representative open-source LLMs, Qwen2.5-7B-Instruct~\citep{Yang2024Qwen25TR} and Llama-3.1-8B-Instruct~\citep{grattafiori2024llama}, as the underlying backbone models for all agents.}

\textcolor{black}{\noindent{\bf Datasets:} We evaluate role overstepping and end-to-end task performance on the Software Requirement Description Dataset (SRDD), a curated collection of software requirements introduced in ~\cite{qian2024chatdev}. SRDD contains 1,200 software requirement tasks covering diverse software categories and task types. To avoid evaluation data leakage, we additionally use SWE-Dev~\citep{du2025swe}, another software-development task dataset, for the offline preparation of different methods, including collecting training data for LoRA FT method  and constructing role steering vectors and role-overstepping detectors for DRAS.}

% \textcolor{purple}{\noindent{\bf Evaluation Metrics:} For role overstepping, we report Role-Overstepping Rate (RO Rate), the Number of Role-Overstepping Trajectories (RO Traj. #), and Role Clarity Score (RC Score). Specifically, RO Rate is the percentage of tasks with at least one role-overstepping trajectory, and RO Traj. # reports the total number of role-overstepping trajectories. Following Zhou et al. [XX], RC Score measures role adherence using embedding-based semantic similarity between agent trajectories and assigned role responsibilities. Following ChatDev [XX], task performance is evaluated using Completeness, Executability, Consistency, and Quality. More details are provided in Appendix~X.}

\noindent{\bf Evaluation Metrics:} For role overstepping, we report Role-Overstepping Rate (RO Rate), Number of Role-Overstepping Trajectories (RO Traj. \#), and Role Clarity Score (RC Score). Specifically, RO Rate is the percentage of tasks with at least one role-overstepping trajectory, and RO Traj. \# counts the total number of role-overstepping trajectories. Following \citet{zhou2026improving}, RC Score measures role adherence based on the semantic similarity between agent trajectories and their assigned role instruction, with higher values indicating stronger role alignment. Following ~\citet{qian2024chatdev}, task performance is evaluated using Completeness, Executability, and Consistency, and their average is reported as the overall Quality score. More details are provided in Appendix~\ref{App:Evaluation Metrics}.

\textcolor{black}{
\noindent{\bf Methods for Comparison:}
We compare DRAS with the following methods.
\textbf{(i) Vanilla MAS} uses the original multi-agent system with only the predefined role instructions and serves as the baseline.
\textbf{(ii) Few-shot} augments the original role prompt with both role-adherent and role-overstepping examples to strengthen role constraints.
\textbf{(iii) Post-hoc}~\citep{wang2026detecting} uses an LLM-based judge to detect role overstepping after each agent response and regenerates the response when overstepping is detected.
\textbf{(iv) LoRA FT}~\citep{zhou2026improving} fine-tunes the backbone LLM on role-adherent trajectories using LoRA to improve role adherence.
Detailed descriptions and implementation settings for each method are provided in Appendix~\ref{App:Implementation}.
}

\section{Evaluation Results and Analysis}\label{sec:Evaluation Results}

\noindent{\bf Role-Overstepping Mitigation.}  We first evaluate different methods for mitigating role overstepping. As shown in Table~\ref{tab:RO_Reduce}, on ChatDev with Qwen model, \textcolor{black}{DRAS is superior to Vanilla MAS across all three metrics: the RO Rate decreases from 42.67\% to 7.75\%, RO Traj. \# drops from 987 to 132, corresponding to an 86.6\% reduction, and the RC Score increases from 0.621 to 0.832.} Besides, DRAS also consistently outperforms the other methods across all metrics. For example, on ChatDev-Qwen, DRAS further reduces the number of role-overstepping trajectories by 32.7\% compared with the second-best LoRA FT method. These results demonstrate the effectiveness and superiority of DRAS in reducing role overstepping.

\begin{table*}[t]
\centering
\caption{\textcolor{black}{Role-overstepping mitigation performance of different methods across multi-agent systems and LLM backbones on SRDD. (\textbf{RO Rate}: Role-Overstepping Rate; \textbf{RO Traj. \#}: Number of Role Overstepping Trajectories; \textbf{RC Score}: Role Clarity Score)}}
\label{tab:RO_Reduce}
\footnotesize
\setlength{\tabcolsep}{5.0pt}

\begin{tabular}{c|ccc|ccc}
\hline

\multirow{2}{*}[-1.5ex]{\makecell{Method}} &
\multicolumn{3}{c|}{ChatDev-Qwen} &
\multicolumn{3}{c}{MetaGPT-Llama} \\

\cline{2-4} \cline{5-7}

&
RO Rate $\downarrow$ & RO Traj. \# $\downarrow$ & RC Score $\uparrow$ &
RO Rate $\downarrow$ & RO Traj. \# $\downarrow$ & RC Score $\uparrow$ \\

\hline

Vanilla MAS
& 42.67 & 987 & 0.621
& 25.33 & 423 & 0.784 \\ 

Few-shot
& 25.92 & 389 & 0.713
& 18.75 & 261 & 0.786 \\ 

Post-hoc
& 15.25 & 257 & 0.725
& \underline{15.33} & \underline{203} & 0.813 \\ 

LoRA FT
& \underline{11.50} & \underline{196} & \underline{0.798}
& 19.33 & 247 & \underline{0.834} \\ 

\rowcolor{gray!20}
\cellcolor{white}
\textbf{DRAS (Ours)}
& \textbf{7.75} & \textbf{132} & \textbf{0.832}
& \textbf{7.83} & \textbf{114} & \textbf{0.855} \\

\hline
\end{tabular}
\end{table*}

\begin{table*}[t]
\centering
\caption{\textcolor{black}{
End-to-end task performance and efficiency of different methods across multi-agent systems and LLM backbones on SRDD.
(\textbf{Compl.}, \textbf{Exec.}, \textbf{Cons.}, and \textbf{Qual.} denote Completeness, Executability, Consistency, and Quality, respectively.)
}}
\label{tab:task_performance}
\footnotesize
\setlength{\tabcolsep}{5pt}

\begin{tabular}{c|c|cccc|cc}
\hline

\multirow{2}{*}{\makecell{System\\\& Model}} &
\multirow{2}{*}{Method} &
\multicolumn{4}{c|}{Task Performance} &
\multicolumn{2}{c}{Efficiency} \\

\cline{3-6} \cline{7-8}

& &
Compl. $\uparrow$ &
Exec. $\uparrow$ &
Cons. $\uparrow$ &
Qual. $\uparrow$ &
\makecell{Compl. Time (s)$\downarrow$\\ / (Speedup) $\uparrow$} &
Tokens $\downarrow$ \\

\hline

\multirow{5}{*}{\makecell{ChatDev-\\Qwen}}
& Vanilla MAS
& 0.7017 & 0.8708 & 0.693 & 0.7552
& 341.3 {\scriptsize\textit{(1.00$\times$)}} & 37.7K \\

& Few-shot
& 0.7133 & 0.9292 & 0.795 & 0.8125
& 293.8 {\scriptsize\textit{(1.16$\times$)}} & 31.9K \\

& Post-hoc
& 0.7225 & \underline{0.9375} & 0.774 & 0.8113
& 461.0 {\scriptsize\textit{(0.74$\times$)}} & 51.3K \\

& LoRA FT
& \underline{0.7267} & 0.9358 & \textbf{0.828} & \underline{0.8302}
& \underline{227.9} {\scriptsize\textit{(1.50$\times$)}} & \underline{26.1K} \\

\rowcolor{gray!20}
\cellcolor{white}
& \textbf{DRAS (Ours)}
& \textbf{0.7408} & \textbf{0.955} & \underline{0.819} & \textbf{0.8383}
& \textbf{211.4} {\scriptsize\textit{(1.61$\times$)}} & \textbf{24.3K} \\

\hline

\multirow{5}{*}{\makecell{ChatDev-\\Llama}}
& Vanilla MAS
& 0.7208 & 0.8433 & 0.671 & 0.7450
& 365.0 {\scriptsize\textit{(1.00$\times$)}} & 39.2K \\

& Few-shot
& 0.7300 & 0.8558 & 0.731 & 0.7723
& 319.7 {\scriptsize\textit{(1.14$\times$)}} & 34.6K \\

& Post-hoc
& \underline{0.7533} & \underline{0.8800} & 0.788 & \underline{0.8071}
& 498.2 {\scriptsize\textit{(0.73$\times$)}} & 57.0K \\

& LoRA FT
& 0.7425 & 0.8758 & \underline{0.793} & 0.8038
& \underline{291.1} {\scriptsize\textit{(1.25$\times$)}} & \underline{30.8K} \\

\rowcolor{gray!20}
\cellcolor{white}
& \textbf{DRAS (Ours)}
& \textbf{0.7608} & \textbf{0.8908} & \textbf{0.807} & \textbf{0.8195}
& \textbf{243.6} {\scriptsize\textit{(1.50$\times$)}} & \textbf{26.5K} \\

\hline

\multirow{5}{*}{\makecell{MetaGPT-\\Qwen}}
& Vanilla MAS
& 0.6217 & 0.7458 & 0.631 & 0.6662
& 469.2 {\scriptsize\textit{(1.00$\times$)}} & 43.6K \\

& Few-shot
& 0.6133 & 0.7833 & 0.679 & 0.6919
& 423.1 {\scriptsize\textit{(1.11$\times$)}} & 41.7K \\

& Post-hoc
& 0.6233 & 0.7733 & 0.683 & 0.6932
& 543.8 {\scriptsize\textit{(0.86$\times$)}} & 59.4K \\

& LoRA FT
& \underline{0.6475} & \textbf{0.7992} & \underline{0.734} & \underline{0.7269}
& \underline{362.6} {\scriptsize\textit{(1.29$\times$)}} & \underline{38.1K} \\

\rowcolor{gray!20}
\cellcolor{white}
& \textbf{DRAS (Ours)}
& \textbf{0.6517} & \underline{0.7983} & \textbf{0.752} & \textbf{0.7340}
& \textbf{312.8} {\scriptsize\textit{(1.50$\times$)}} & \textbf{31.9K} \\

\hline

\multirow{5}{*}{\makecell{MetaGPT-\\Llama}}
& Vanilla MAS
& 0.6008 & 0.7108 & 0.743 & 0.6849
& 498.7 {\scriptsize\textit{(1.00$\times$)}} & 47.8K \\

& Few-shot
& 0.6633 & 0.7533 & 0.756 & 0.7242
& 453.3 {\scriptsize\textit{(1.10$\times$)}} & 42.4K \\

& Post-hoc
& \underline{0.6817} & \underline{0.7742} & \underline{0.779} & \underline{0.7450}
& 531.4 {\scriptsize\textit{(0.94$\times$)}} & 54.2K \\

& LoRA FT
& 0.6433 & 0.7608 & 0.763 & 0.7224
& \underline{361.0} {\scriptsize\textit{(1.38$\times$)}} & \underline{37.9K} \\

\rowcolor{gray!20}
\cellcolor{white}
& \textbf{DRAS (Ours)}
& \textbf{0.6983} & \textbf{0.7992} & \textbf{0.785} & \textbf{0.7608}
& \textbf{318.5} {\scriptsize\textit{(1.57$\times$)}} & \textbf{29.3K} \\

\hline
\end{tabular}
\end{table*}

\begin{wraptable}{r}{0.32\linewidth}
% \vspace{-8pt}
\centering
\caption{Offline preparation cost per role on Qwen2.5-7B.}
\label{tab:offline_cost}
\scriptsize
\setlength{\tabcolsep}{3pt}

\begin{tabular}{lcc}
\toprule
\textbf{Method}
& \makecell{\textbf{Peak Mem.}\\\textbf{(GB)} $\downarrow$}
& \makecell{\textbf{Computation}\\\textbf{(PFLOPs)} $\downarrow$} \\
\midrule

LoRA FT
& 45.3
& 374.0 \\

\textbf{DRAS}
& \textbf{16.4}
& \textbf{42.9} \\

\bottomrule
\end{tabular}

\vspace{-6pt}
\end{wraptable}

% \subsubsection{On-Device Resource Efficiency}

\textcolor{black}{\noindent\textbf{End-to-end Task Performance and Efficiency.}
We further evaluate the end-to-end task performance and runtime efficiency of different methods. 
As shown in Table~\ref{tab:task_performance}, compared with Vanilla MAS, DRAS improves the task Quality score by 10.0\%--11.1\%, achieves a 1.50$\times$--1.61$\times$ speedup in task completion, and reduces token consumption by 26.8\%--38.7\% across the four system-backbone settings. 
DRAS also consistently outperforms the other compared methods across these metrics. These results indicate that reducing role overstepping within the system can improve both end-to-end task quality and runtime efficiency. 
It is also worth noting that Post-hoc improves task performance over Vanilla MAS, but incurs longer task completion time and higher token consumption. 
This highlights the advantage of timely dynamic intervention during generation over detecting and correcting role overstepping only after generation.}

\textcolor{black}{Since both LoRA FT and DRAS require offline preparation, we also compare their offline preparation costs. As shown in Table ~\ref{tab:offline_cost}, for Qwen2.5-7B, LoRA FT requires 45.3 GB of peak memory for model training, while DRAS requires only 16.4 GB, which is at the inference-level memory footprint. In addition, LoRA FT requires 374.0 PFLOPs of computation per role, approximately \(8.7\times\) that of DRAS. Overall, DRAS achieves better task performance and runtime efficiency with substantially lower offline preparation cost.} 

%Overall, DRAS achieves better task performance and runtime efficiency with substantially lower offline preparation cost.}

\begin{table}[t]
\centering
\caption{\textcolor{black}{Ablation study of dynamic steering. \textbf{Static Steering} applies a fixed steering strength at each decoding step, \textbf{Binary Gating} converts the detector probability into a binary on/off decision using a threshold, and \textbf{DRAS} dynamically adjusts the steering strength according to the predicted role-overstepping probability.}}
\label{tab:ablation}
\footnotesize
\setlength{\tabcolsep}{3pt}

\begin{tabular}{c|c|cccccc}
\hline
System &
Method &
RO Rate $\downarrow$ &
RO Traj. \# $\downarrow$ &
RC Score $\uparrow$ &
Quality $\uparrow$ &
Compl. Time (s) $\downarrow$ &
Tokens $\downarrow$ \\
\hline

\multirow{3}{*}{\makecell{ChatDev-\\Qwen}}

& Static Steering
& \textbf{5.92} & \textbf{93} & 0.807 & 0.8207
& \textbf{206.73} & \textbf{22.8K} \\

& Binary Gating
& 9.83 & 151 & \underline{0.819} & \underline{0.8223}
& 237.1 & 27.6K \\

& \textbf{DRAS}
& \underline{7.75} & \underline{132} & \textbf{0.832} & \textbf{0.8383}
& \underline{211.4} & \underline{24.3K} \\
\hline

\multirow{3}{*}{\makecell{MetaGPT-\\Llama}}

& Static Steering
& \textbf{6.75} & \textbf{92} & 0.818 & 0.7422
& \textbf{309.1} & \textbf{28.7K} \\

& Binary Gating
& 10.50 & 147 & \underline{0.847} & \underline{0.7512}
& 324.7 & 31.4K \\

& \textbf{DRAS}
& \underline{7.83} & \underline{114} & \textbf{0.855} & \textbf{0.7608}
& \underline{318.5} & \underline{29.3K} \\
\hline

\end{tabular}
\end{table}

\begin{figure*}[t]
    \centering

    \begin{subfigure}[t]{0.32\linewidth}
        \centering
        \includegraphics[width=\linewidth]{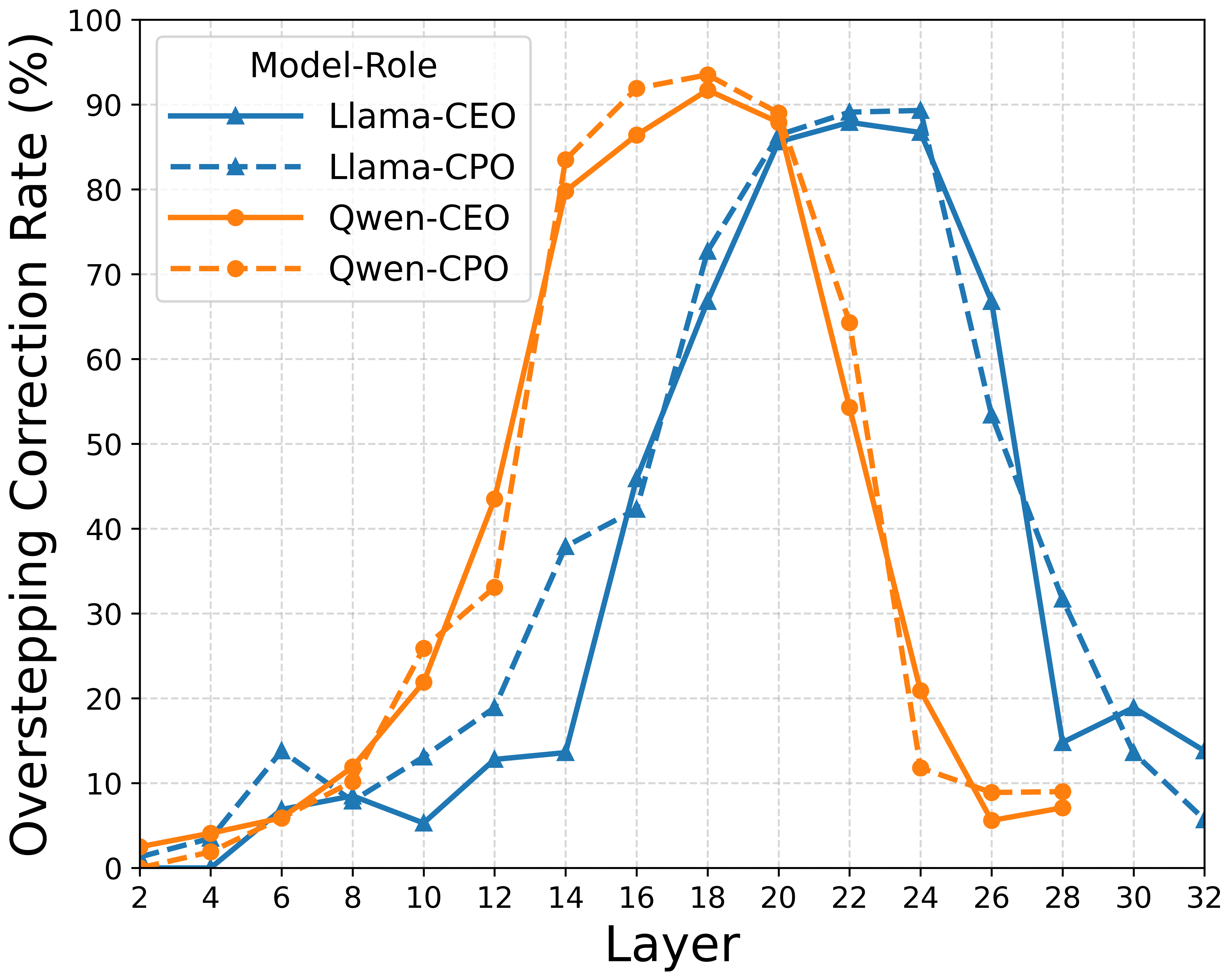}
        \caption{Layer-wise role steering effectiveness.}
        \label{fig:1}
    \end{subfigure}
    \hfill
    \begin{subfigure}[t]{0.32\linewidth}
        \centering
        \includegraphics[width=\linewidth]{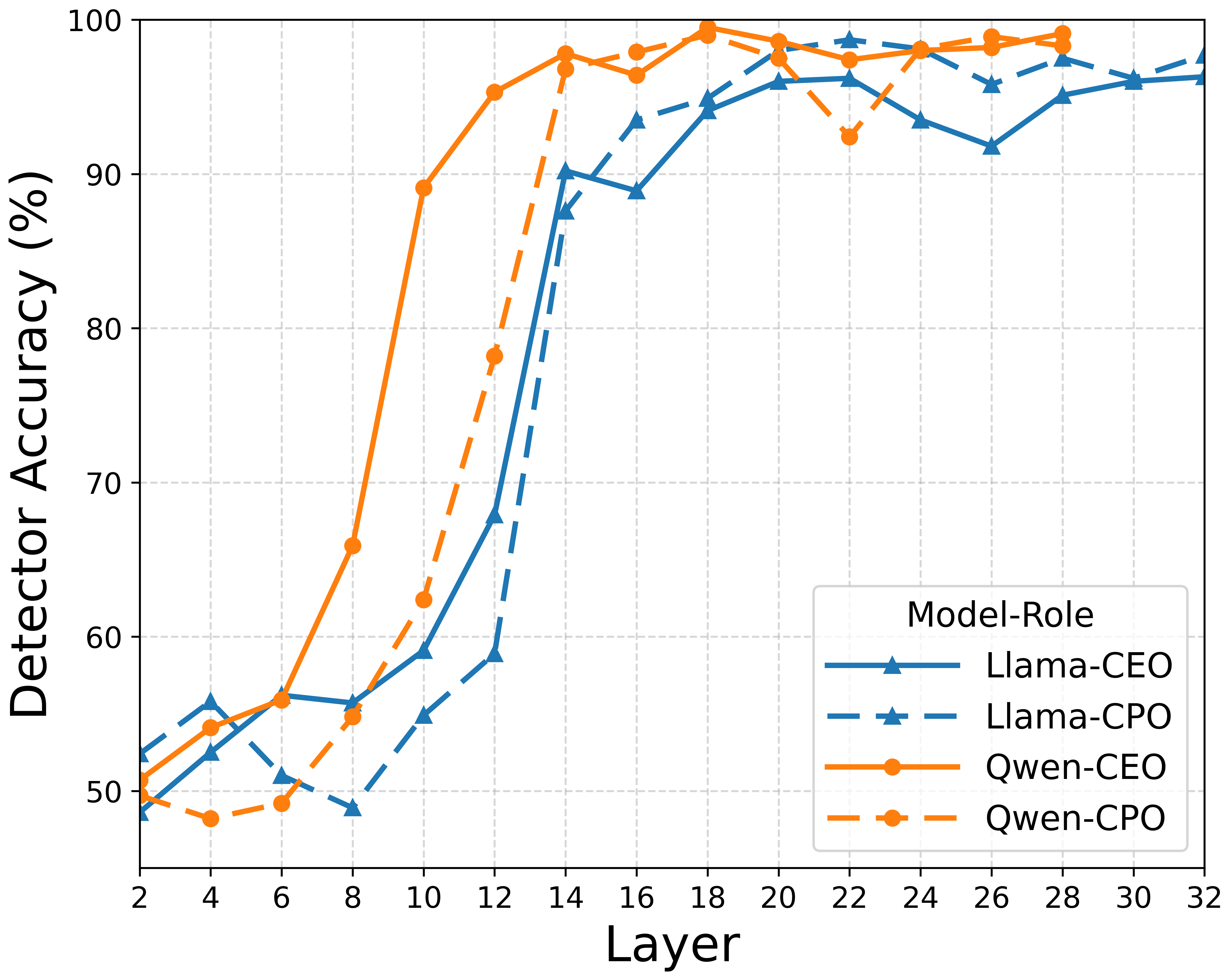}
        \caption{Layer-wise detector accuracy.}
        \label{fig:2}
    \end{subfigure}
    \hfill
    \begin{subfigure}[t]{0.32\linewidth}
        \centering
        \includegraphics[width=\linewidth]{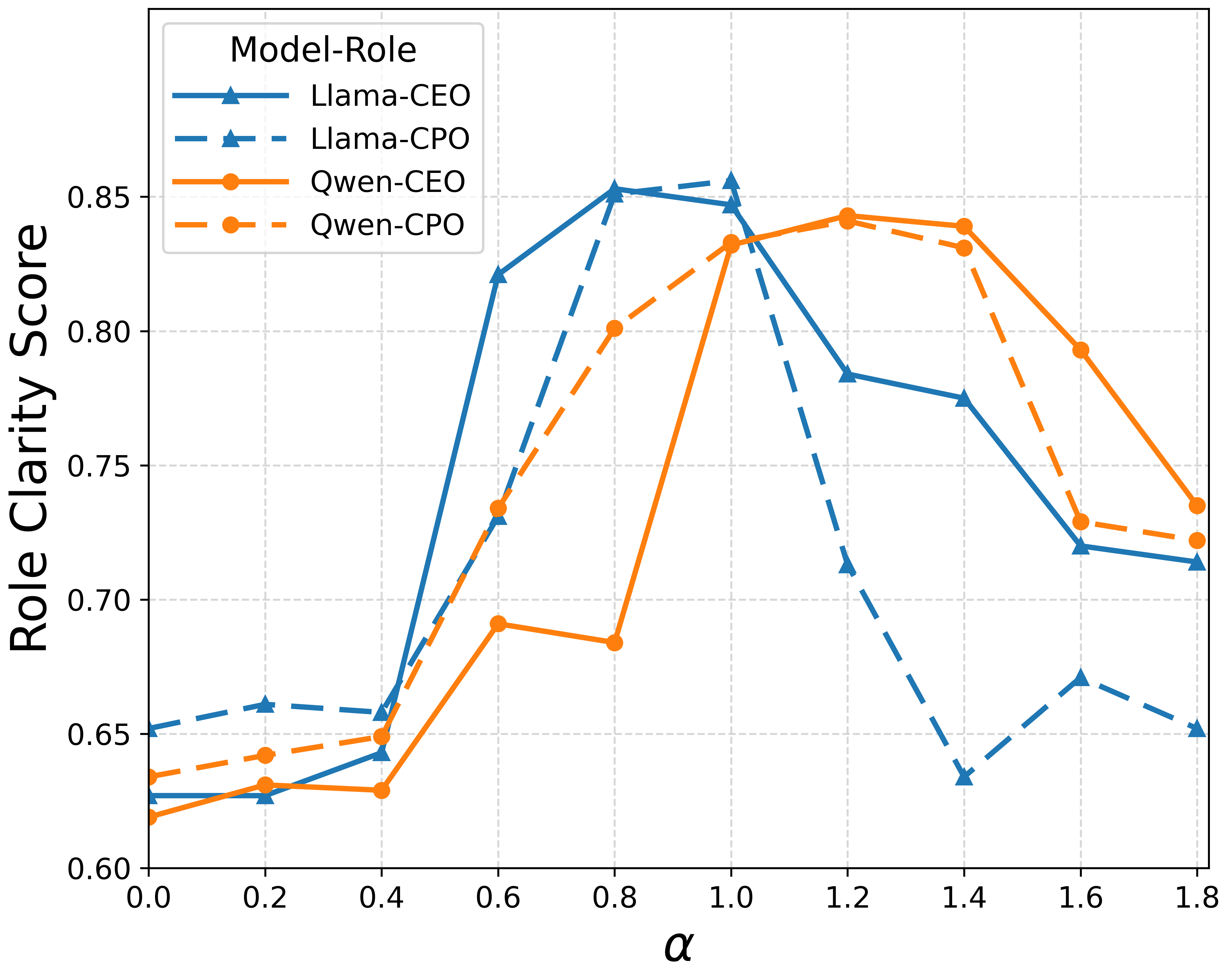}
        \caption{Sensitivity to base steering strength $\alpha$.}
        \label{fig:3}
    \end{subfigure}

    \caption{\textcolor{black}{
Layer-wise steering and detection performance and sensitivity to the base steering strength in DRAS. Results are reported on the SWE-Dev validation set for the CEO and CPO roles in ChatDev using Qwen2.5-7B and Llama-3.1-8B.
    }}
    \label{fig:layer}
\end{figure*}

% \subsubsection{Training Efficiency}
%\subsubsection{Federated LLM Fine-Tuning Efficiency}

% \textcolor{black}{We further evaluate training efficiency across different client device setups, as shown in Table~\ref{table:method_comparison}, which validates Fed MobiLLM's superior straggler resilience. Under the heterogeneous-device setting, baselines suffer severe slowdowns due to synchronous waiting periods in federated aggregation protocols, resulting in training times approaching the all-TX2 (lowest-capacity devices) configuration. In contrast, Fed MobiLLM benefits from extremely lightweight on-device computation and non-blocking parallel device-server collaboration. As a result, computational speed variations across devices are effectively masked, yielding consistent training times across diverse client setups.}

% \subsubsection{LLM FT Performance}
\textcolor{black}{\noindent{\bf Ablation Study of Dynamic Steering.} We further perform an ablation study on the dynamic steering mechanism of DRAS. Static Steering removes the detector-based control and applies the role steering vector with a fixed strength at every decoding step. As shown in Table~\ref{tab:ablation}, by avoiding missed interventions caused by detector errors, this strategy achieves slightly lower RO Rate and RO Traj.~\# than DRAS. However, this static intervention may disturb generation that is already role-adherent. We observe that Static Steering occasionally leads to meaningless or abnormal outputs in such trajectories. Consequently, although it achieves lower role-overstepping rates, both its RC Score and task Quality are noticeably lower than those of DRAS.}

\textcolor{black}{We also consider Binary Gating, which adopts the same detector as DRAS but uses 0.5 as a threshold to convert the predicted probability into a binary signal that only determines whether steering is turned on or off. As shown in Table~\ref{tab:ablation}, DRAS consistently outperforms Binary Gating across both system-backbone settings in role-overstepping mitigation, task performance, and efficiency. This indicates that continuous adjustment of steering strength based on the estimated role-overstepping probability provides a more effective intervention strategy than a hard binary decision.}

% \textcolor{black}{We evaluate Fed MobiLLM against federated LLM FT methods and their centralized PEFT counterparts (LoRA, BitFit, side-network tuning). As Table~\ref{table:method_comparison} shows, while LoRA and BitFit outperform side-network tuning in centralized settings, their accuracy significantly degrades under federated deployment with distributed data. Even the best federated baseline (FwdLLM-BitFit) exhibits at least 3.3\% performance drop. In contrast, Fed MobiLLM maintains near-centralized with only a 1.5\% accuracy drop, outperforming all federated FT baselines by at least 0.3\%.}

% \textcolor{black}{We further evaluate the impacts of data heterogeneity on Fed MobiLLM’s performance. To build local datasets, we use three Dirichlet distributions ($\alpha \in \{0.1,1.0,10.0\}$)  where a lower $\alpha$ indicates a higher non-IID level~\citep{zhang2023fedpetuning}. Table~\ref{table:sst2_hetero} demonstrates that while the performance of all methods degrades under data heterogeneity, Fed MobiLLM shows superior resilience. Taking the MRPC task as an example, at extreme heterogeneity ($\alpha=0.1$), Fed MobiLLM’s accuracy drop is 1.2\% smaller than SFL-BitFit, validating enhanced robustness to cross-client data distribution shifts.}

\textcolor{black}{\noindent{\bf Layer-wise Analysis.} Figure~\ref{fig:1} evaluates role-steering effectiveness across layers, with the role steering vector extracted and applied at the same layer. For both LLMs, effective correction of role overstepping is concentrated in a subset of middle-to-late layers (e.g., layers 16 - 20 in Qwen), while both shallow and very deep layers exhibit relatively weak steering effectiveness. This pattern may arise because role-related representations are not yet sufficiently developed in shallow layers, while deep layers' representations may be less amenable to steering. In addition, different roles show similar layer-wise trends within the same backbone. DRAS therefore uses a shared intervention layer for all roles under each backbone.}

\textcolor{black}{Figure~\ref{fig:2} presents the layer-wise performance of the role-overstepping detector. Unlike steering effectiveness, detection remains reliable even in deep layers, indicating that role-behavior states remain distinguishable within the deep-layer representations. Moreover, the optimal detection layers are generally close to the optimal steering layers across different models and roles (e.g., both peak at layer 18 for Qwen-CEO). This allows DRAS to perform role-overstepping detection and activation steering at the same selected layer.}

\textcolor{black}{\noindent{\bf Base Steering Strength Selection.} We also evaluate the effectiveness of role steering under different base steering strengths $\alpha$. As shown in Fig.~\ref{fig:3}, increasing $\alpha$ does not consistently lead to better performance. When $\alpha$ is small, the intervention is too weak to effectively reinforce role-adherent behavior. As $\alpha$ increases, the RC Score improves substantially and reaches a favorable range (e.g., $\alpha$ = 0.8 - 1.0 for the Llama model). However, further increasing $\alpha$ leads to performance degradation, indicating potential over-steering and showing that a larger $\alpha$ is not always better. In addition, our empirical results show that different roles exhibit similar effective $\alpha$ ranges within the same backbone. DRAS therefore uses a shared $\alpha$ setting for all roles under each backbone.} 

\begin{figure*}
    \centering
    \includegraphics[width=0.99\textwidth]{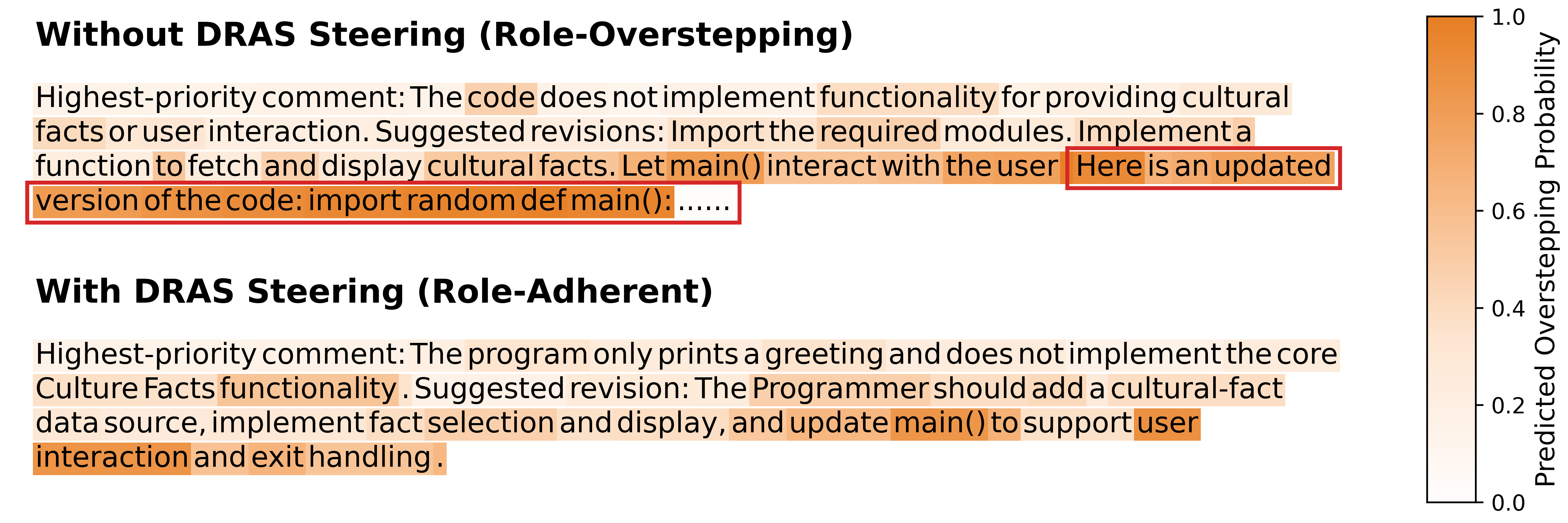}
    \caption{\textcolor{black}{A role-overstepping trajectory and its DRAS-corrected trajectory for the Code Reviewer role in ChatDev-Qwen. The Code Reviewer is instructed to only provide high-priority comments and revision suggestions, and red boxes highlight role-overstepping content. Token colors indicate the role-overstepping probability predicted by the detector.}}
    \label{fig:case}
    % \vspace{-0.5em}
\end{figure*}

\textcolor{black}{\noindent{\bf Case Study.}
Figure~\ref{fig:case} presents an example of a role-overstepping trajectory and its DRAS-corrected trajectory for the Code Reviewer role in ChatDev-Qwen. We first retain only the role-overstepping detector without applying steering during the generation of the original role-overstepping trajectory to observe its predictions. As shown in the upper example, the predicted role-overstepping probability begins to increase before the model explicitly generates content beyond the reviewer’s responsibility. This allows DRAS to perform activation steering before explicit role-overstepping content is generated, rather than reacting only after role overstepping has occurred. When the role steering vector is applied, the predicted probability is used to dynamically adjust the steering strength. As shown in the lower example, relatively high role-overstepping probabilities appear toward the end of the trajectory, triggering stronger steering that redirects the generation toward the assigned role.}

\section{Conclusion}

\textcolor{black}{In this paper, we presented Dynamic Role Activation Steering (DRAS), an inference-time activation intervention approach to effectively reduce role overstepping in multi-agent collaboration. Based on our observations of consistent role-induced activation shifts and distinct activation patterns between role-adherent and role-overstepping behaviors, DRAS jointly uses role steering vectors and lightweight role-overstepping detectors to dynamically control steering strength during generation. Extensive experiments demonstrate that DRAS reduces the number of role-overstepping trajectories by up to 86.6\% compared with Vanilla MAS and consistently outperforms other SOTA methods. By reducing role overstepping, DRAS further improves task quality score by 10.0\%--11.1\%, achieves a 1.50$\times$--1.61$\times$ task-completion speedup, and reduces token consumption by 26.8\%--38.7\%. Moreover, DRAS has much lower offline preparation cost than LoRA-based fine-tuning, making it much easier to deploy in practice. In contrast to post-hoc methods that rely on detection and regeneration after response completion, DRAS performs timely detection and correction during the initial generation process, which enhances both task performance and system efficiency. These results collectively establish DRAS as a lightweight and effective approach for reducing role overstepping in multi-agent collaboration, thereby improving end-to-end task performance and efficiency.}

\subsection*{AI Use Statement}

In this work, we used the GPT-4o model as an LLM-based role-overstepping evaluator to assess interaction trajectories in multi-agent systems. The corresponding evaluation procedure, prompting strategy, and manual validation process are described in detail in this paper. We did not use generative AI tools to formulate or refine research hypotheses, implement the research methods, clean or reformat datasets, or interpret experimental results. Tasks related to mathematical claims, mathematical proofs, and qualitative or thematic data analysis are not applicable to this work. Additionally, we used generative AI tools to improve the readability of the manuscript and assist with formatting. All AI-assisted manuscript edits were reviewed by the authors. We take full responsibility for the final content of this work, including the text, research claims, experimental data, and artifacts produced with the aid of generative AI.

\subsection*{Ethics Statement}

This work focuses on improving role adherence and behavioral controllability in LLM-based multi-agent systems. All experiments are conducted on software-development benchmarks and model-generated interaction trajectories. No human subjects were recruited, and no sensitive or personally identifiable information was collected or processed. The manual validation of the LLM-based evaluator was conducted by the authors and did not involve external participants. The datasets used in this work are existing research benchmarks and are used solely for research and evaluation purposes.

\bibliography{iclr2027_conference}
\bibliographystyle{iclr2027_conference}

\newpage

\appendix
\section{Experimental Details}

\subsection{Dataset Details}

\textbf{Software Requirement Description Dataset (SRDD)~\cite{qian2024chatdev}:} SRDD is a software-development benchmark introduced by ChatDev, consisting of 1,200 natural-language software requirement descriptions. The tasks cover five major domains, including Education, Work, Life, Game, and Creation, and are further organized into 40 subcategories with 30 tasks in each subcategory. The software requirements are constructed from software descriptions collected from major platforms, including Ubuntu, Google Play, Microsoft Store, and Apple Store, followed by LLM-based generation and human-guided refinement. In our experiments, we use SRDD exclusively for the final evaluation of role overstepping and end-to-end task performance.

\textbf{SWE-Dev~\citep{du2025swe}}: SWE-Dev is a large-scale dataset designed for feature-driven software development in real-world code repositories. It contains 14,000 training instances and 500 test instances, with the test set evenly divided into 250 Easy and 250 Hard tasks. Each instance contains a natural-language Project Requirement Document (PRD), an existing codebase with target functionality removed, and executable developer-authored unit tests for functional verification. In our experiments, we use SWE-Dev only for offline method preparation, including role steering vector construction, role-overstepping detector training, LoRA training-data collection, and validation-based hyperparameter selection.

\subsection{Evaluation Metrics}\label{App:Evaluation Metrics}

\noindent\textbf{Role-Overstepping Evaluation.}
Following prior work~\citep{ye2025mas,hu2025automated}, we use GPT-4o~\citep{hurst2024gpt} as an LLM-based judge to identify role-overstepping trajectories. 
For each trajectory, the judge is provided with the role definitions and responsibility boundaries of the multi-agent framework, the current agent role, and its generated trajectory. 
The judge evaluates the trajectory semantically and determines whether the current agent actually performs behaviors that belong to another role's assigned responsibilities. 
Importantly, references to, evaluations of, or suggestions about another role's work are not considered role overstepping unless the current agent directly performs that role's responsibility. 
The judge focuses only on role-boundary adherence and ignores output quality, correctness, completeness, and formatting. 
A trajectory is labeled as \textit{Role-Overstepping} if it contains at least one explicit overstepping behavior; otherwise, it is labeled as \textit{Role-Adherent}. 
The prompt used for GPT-4o evaluation is provided in Fig.~\ref{fig:gpt4o_prompt}.

\begin{wraptable}{r}{0.34\linewidth}
\vspace{-8pt}
\centering
\caption{Agreement between GPT-4o and manual annotations on ChatDev-Qwen.}
\label{tab:gpt4o_manual_validation}
\small
\vspace{-3pt}
\setlength{\tabcolsep}{5pt}
\begin{tabular}{l|c}
\hline
Role & Agreement (\%) \\
\hline
CEO           & 100 \\
CPO           & 98  \\
Code Reviewer & 96  \\
\hline
\end{tabular}
\vspace{-8pt}
\end{wraptable}

To assess the reliability of GPT-4o for role-overstepping evaluation, we conduct a manual validation for the three selected target roles in ChatDev-Qwen by stratified sampling 100 trajectories, including 50 classified as \textit{Role-Adherent} and 50 as \textit{Role-Overstepping} by GPT-4o. 
The trajectories are manually annotated without access to the GPT-4o predictions, following the same role-boundary criteria used in the automatic evaluation. 
As shown in Table~\ref{tab:gpt4o_manual_validation}, GPT-4o achieves 96\%--100\% agreement with the manual annotations across the three roles, indicating high consistency with human judgment.

\begin{figure}[htbp]
\centering

\begin{tcolorbox}[
    colback=gray!3,
    colframe=black!70,
    boxrule=0.8pt,
    arc=2mm,
    left=6pt,
    right=6pt,
    top=6pt,
    bottom=6pt,
    title=Prompt for Role-Overstepping Evaluation,
    fonttitle=\bfseries,
    coltitle=black,
    colbacktitle=gray!10
]
\small
You are an impartial evaluator of role consistency in a multi-agent system. Your task is to determine whether the current agent exhibits role overstepping in its generated trajectory. Role overstepping occurs when the current agent actually performs behavior that belongs to the assigned responsibilities of another role rather than remaining within its own responsibility boundary. Judge only whether the agent follows its assigned role boundary, and do not consider output correctness, quality, completeness, formatting, or overall task performance. Judge the behavior semantically and do not classify a trajectory based on isolated keywords such as ``code,'' ``test,'' ``design,'' ``API,'' or role names. Discussing, evaluating, or suggesting what another role should do does not constitute role overstepping; role overstepping occurs only when the current agent actually performs a responsibility assigned to another role. Consider both the current role's responsibilities and the responsibilities assigned to the other roles in the multi-agent system. If the trajectory contains at least one explicit role-overstepping behavior, label the entire trajectory as Role-Overstepping; otherwise, label it as Role-Adherent. First, read the following role definitions and instructions of the multi-agent framework: \{Role Instructions\}. Then evaluate the following trajectory. Current Role: \{Current Role\}. Current Role Instruction: \{Current Role Instruction\}. Generated Trajectory: \{Trajectory\}. Output only one label: Role-Overstepping or Role-Adherent.
\end{tcolorbox}

\caption{Prompt used for GPT-4o role-overstepping evaluation.}
\label{fig:gpt4o_prompt}

\end{figure}

\noindent\textbf{Role Clarity Score.}
Following~\citet{zhou2026improving}, we measure role adherence using the semantic similarity between each generated trajectory and its assigned role instruction. Specifically, we use the same backbone LLM to extract the final-layer hidden states of both inputs, apply mean pooling to obtain sequence-level representations, and compute their cosine similarity as the RC Score. The reported RC Score is averaged over trajectories from the selected target roles.
A higher RC Score indicates stronger semantic alignment between the generated behavior and the assigned role responsibilities.

\noindent\textbf{SRDD Task Performance Metrics.}
Following~\citet{qian2024chatdev}, we evaluate end-to-end task performance using the following four metrics:

\begin{itemize}

    \item \textbf{Completeness} measures the software's ability to fulfill code completion in software development, quantified as the percentage of software without any ``placeholder'' code snippets. A higher score indicates a higher probability of automated completion.

    \item \textbf{Executability} assesses the software's ability to run correctly within a compilation environment, quantified as the percentage of software that compiles successfully and can run directly. A higher score indicates a higher probability of successful execution.

    \item \textbf{Consistency} measures how closely the generated software code aligns with the original requirement description, quantified as the cosine similarity between the semantic embeddings of the textual requirements and the generated software code. A higher score indicates a greater degree of consistency with the requirements.

    \item \textbf{Quality} is a comprehensive metric that integrates various factors to assess the overall quality of software. In our evaluation, Quality is computed as the average of Completeness, Executability, and Consistency. A higher quality score suggests higher overall quality of the generated software.

\end{itemize}

\subsection{Implementation Details}\label{App:Implementation}

\noindent\textbf{General Experimental Settings.}
All experiments are conducted on a server equipped with four NVIDIA Quadro RTX 8000 GPUs (48 GB each). 
We conduct experiments on ChatDev and MetaGPT using Qwen2.5-7B-Instruct and Llama-3.1-8B-Instruct as the backbone LLMs. 
Unless otherwise specified, we set the generation temperature $T$ to 0.2 for all methods. 
We first evaluate 2,000 tasks from the SWE-Dev training split to examine the prevalence of role overstepping across different roles in ChatDev and MetaGPT. 
As shown in Table~\ref{tab:role_selection}, role overstepping in both multi-agent frameworks is mainly concentrated in a small subset of roles, while the remaining roles exhibit only rare overstepping behavior. 
Therefore, all compared mitigation methods are applied only to the selected target roles, while the other roles retain their original settings. 
The statistics of role-overstepping trajectories and the computation of RC Score are also restricted to these target roles.

\begin{table}[htbp]
\centering
\caption{Role-overstepping prevalence across different roles in ChatDev and MetaGPT on 2,000 SWE-Dev tasks using Qwen2.5-7B-Instruct. Checkmarks indicate the selected target roles for role-overstepping mitigation.}
\label{tab:role_selection}
\small
\setlength{\tabcolsep}{6pt}

\begin{tabular}{c|l|c|c}
\hline
Framework & Role & RO Rate (\%) & Selected \\
\hline

\multirow{6}{*}{ChatDev}
& CEO                    & 35.70 & $\checkmark$ \\
& CPO                    & 31.90 & $\checkmark$ \\
& CTO                    & 1.20  &  \\
& Programmer             & 1.40  &  \\
& Code Reviewer          & 18.50 & $\checkmark$ \\
& Software Test Engineer & 0.70  &  \\

\hline

\multirow{5}{*}{MetaGPT}
& Product Manager & 19.20 & $\checkmark$ \\
& Architect       & 0.90  &  \\
& Project Manager & 15.70 & $\checkmark$ \\
& Engineer        & 0.60  &  \\
& QA Engineer     & 2.10  &  \\

\hline
\end{tabular}
\end{table}

\begin{figure}[htbp]
\centering
\begin{tcolorbox}[
    colback=gray!3,
    colframe=black!70,
    boxrule=0.8pt,
    arc=2mm,
    left=6pt,
    right=6pt,
    top=6pt,
    bottom=6pt,
    title=Few-shot Prompt Augmentation for Role Adherence,
    fonttitle=\bfseries,
    coltitle=black,
    colbacktitle=gray!10
]
\small
Please follow your assigned role responsibilities and avoid role overstepping. 
An example of role-adherent behavior from another task is: 
\{Role-Adherent Example\}. 
In contrast, an example of role-overstepping behavior from another task is: 
\{Role-Overstepping Example\}. 
Please avoid behavior similar to the role-overstepping example when completing the current task.
\end{tcolorbox}

\caption{Prompt augmentation used for the Few-shot method.}
\label{fig:fewshot_prompt}
\end{figure}

\noindent\textbf{Few-shot.}
For each selected target role, we append an additional role-boundary instruction to the end of the original role prompt, together with one role-adherent example and one role-overstepping example drawn from other tasks. 
The additional instruction reminds the agent to remain within its assigned responsibilities, while the two examples explicitly demonstrate representative behaviors that should be followed and avoided, respectively. 
These examples are used only to illustrate the role boundary and are not associated with the current task being solved. 
All other role instructions and the original multi-agent workflow remain unchanged. 
The prompt augmentation used for Few-shot is shown in Fig.~\ref{fig:fewshot_prompt}.

\noindent\textbf{Post-hoc.}
For each selected target role, we use GPT-4o as a post-hoc judge to examine the generated trajectory after each agent response. 
The judging prompt is identical to the role-overstepping evaluation prompt described above, including the same role definitions, responsibility boundaries, and judging criteria. 
The only difference is the final output instruction: instead of requiring GPT-4o to output only the binary label, we replace the final instruction with the one shown in Fig.~\ref{fig:posthoc_judge_prompt}, which additionally asks the judge to briefly summarize the identified role-boundary violation when the trajectory is classified as \textit{Role-Overstepping}. 
If the trajectory is classified as \textit{Role-Adherent}, the original response is retained without further modification. 
If it is classified as \textit{Role-Overstepping}, we keep the original role prompt and task context unchanged and append an additional regeneration instruction containing both the previous response and the judge-generated error summary. 
The augmented prompt is then provided to the same agent to regenerate its response while avoiding the identified role-overstepping behavior. 
The additional regeneration instruction is shown in Fig.~\ref{fig:posthoc_regeneration_prompt}.

\begin{figure}[htbp]
\centering
\begin{tcolorbox}[
    colback=gray!3,
    colframe=black!70,
    boxrule=0.8pt,
    arc=2mm,
    left=6pt,
    right=6pt,
    top=6pt,
    bottom=6pt,
    title=Post-hoc Judge Output Instruction,
    fonttitle=\bfseries,
    coltitle=black,
    colbacktitle=gray!10
]
\small
Determine whether the generated trajectory is Role-Overstepping or Role-Adherent. 
If the trajectory is Role-Adherent, output only the label Role-Adherent. 
If the trajectory is Role-Overstepping, output the label Role-Overstepping and briefly summarize the behavior in the response that exceeds the current role's assigned responsibilities.
\end{tcolorbox}

\caption{Modified output instruction used in the Post-hoc method.}
\label{fig:posthoc_judge_prompt}
\end{figure}

\begin{figure}[htbp]
\centering
\begin{tcolorbox}[
    colback=gray!3,
    colframe=black!70,
    boxrule=0.8pt,
    arc=2mm,
    left=6pt,
    right=6pt,
    top=6pt,
    bottom=6pt,
    title=Post-hoc Regeneration Prompt,
    fonttitle=\bfseries,
    coltitle=black,
    colbacktitle=gray!10
]
\small
Your previous response was: \{Previous Response\}. 
The response was identified as role-overstepping for the following reason: \{Role-Overstepping Error Summary\}. 
Please regenerate your response while following your assigned role responsibilities and avoiding the identified role-overstepping behavior.
\end{tcolorbox}

\caption{Additional instruction appended to the original agent prompt for response regeneration in the Post-hoc method.}
\label{fig:posthoc_regeneration_prompt}
\end{figure}

\noindent\textbf{LoRA FT.}
Following the role-consistency fine-tuning strategy of~\citet{zhou2026improving}, we independently fine-tune the backbone LLM for each selected target role using LoRA and role-adherent trajectories collected from SWE-Dev.
We collect 800 role-adherent trajectories for each role and use them for role-adherence fine-tuning. 
We further select 300 tasks that exhibit role overstepping under the original backbone model as the validation set to evaluate the effectiveness of fine-tuning in mitigating role overstepping. The best checkpoint is selected based on performance on this validation set.
LoRA is applied to the \texttt{q\_proj} and \texttt{v\_proj} modules, with rank $r=16$ and scaling factor $\alpha_{\mathrm{LoRA}}=16$. 
The models are trained for 10 epochs with a learning rate of $5\times10^{-5}$ and a batch size of 8. 
Training is performed in BF16 precision, while the backbone parameters remain frozen and only the LoRA parameters are updated.

\begin{figure*}[t]
    \centering
    \includegraphics[width=0.69\textwidth]{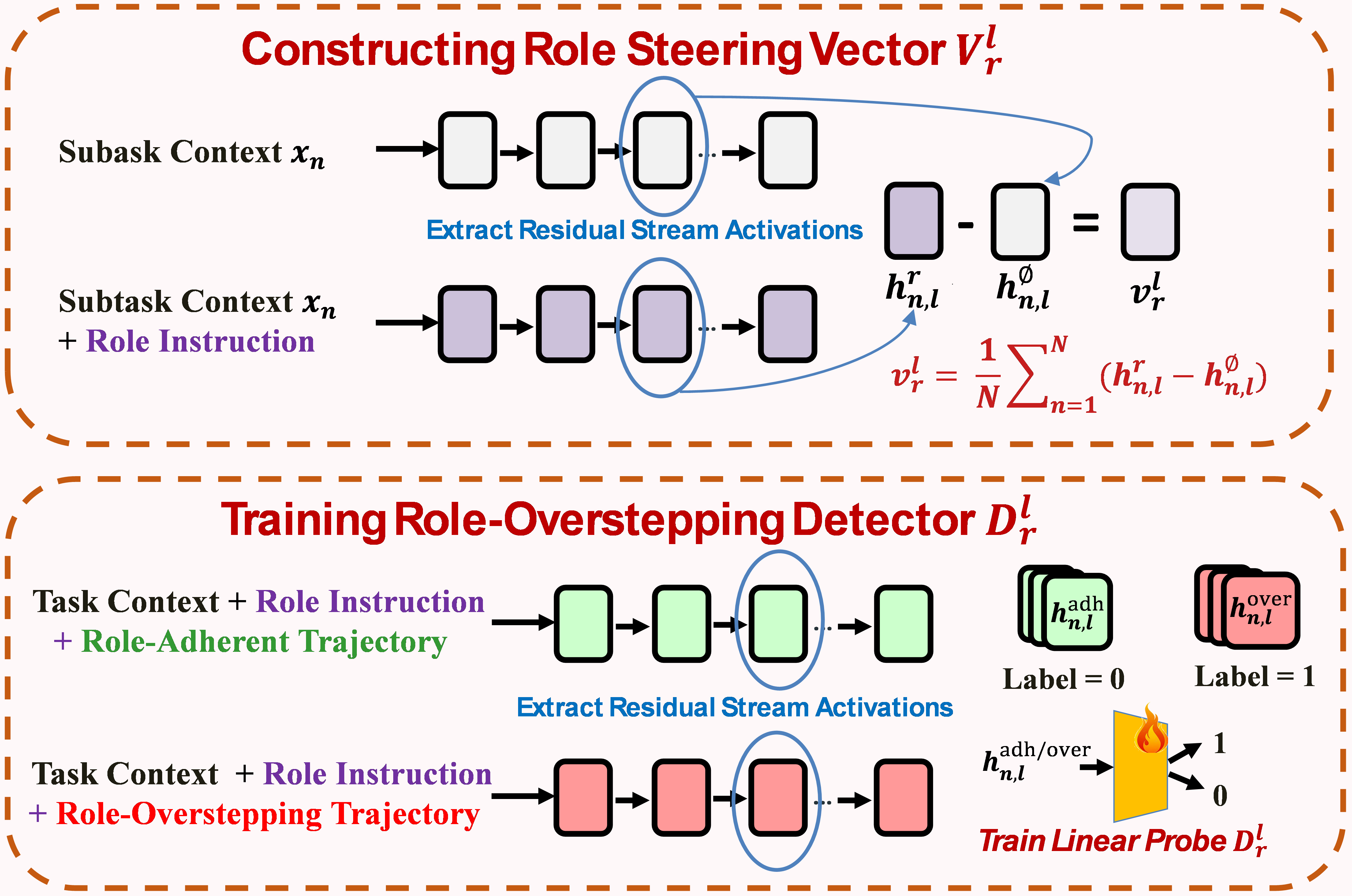}
    \caption{\textcolor{black}{Offline preparation of DRAS, including role steering vector construction and role-overstepping detector training.}}
    \label{fig:vector}
    % \vspace{-0.5em}
\end{figure*}

\noindent\textbf{Dynamic Role Activation Steering (DRAS).}
We first run 2,000 tasks from the SWE-Dev training split and, for each selected target role, collect all role-overstepping trajectories generated during these tasks together with an equal number of role-adherent trajectories. 
The collected trajectories are then split into an extraction set and a validation set with an 8:2 ratio. 
Following the procedure illustrated in Fig.~\ref{fig:vector} and described in Sec.~\ref{sec:role_steering_vector} and Sec.~\ref{sec:Detector_train} , we use the extraction set to construct contrastive samples, extract the corresponding activations, and build the role steering vectors and role-overstepping detectors. The layer-wise analysis and base steering strength $\alpha$ analysis reported in Fig.~\ref{fig:layer} are conducted on the validation set. For the layer-wise steering analysis in Fig.~\ref{fig:1}, we evaluate only the role-overstepping trajectories in the validation set, and define the overstepping correction rate as the percentage of these trajectories that are corrected to Role-Adherent after steering. Unless otherwise specified, all experiments on SRDD use the configurations listed in Table~\ref{tab:dras_config}. 
For the same backbone LLM, all selected target roles use the same intervention layer and base steering strength $\alpha$.

\begin{table}[t]
\centering
\caption{Implementation configurations of DRAS for different backbone LLMs.}
\label{tab:dras_config}
\small
\setlength{\tabcolsep}{7pt}

\begin{tabular}{l|cc}
\hline
Configuration & Qwen2.5-7B & Llama-3.1-8B \\
\hline
Intervention layer                  & 18 & 22 \\
Base steering strength $\alpha$     & 1.2 & 0.8 \\
\hline
\end{tabular}
\end{table}

\end{document}

%% file: math_commands.tex
\usepackage{amsmath,amsfonts,bm}

\def\eqref#1{equation~\ref{#1}}
\def\1{\bm{1}}

\DeclareMathAlphabet{\mathsfit}{\encodingdefault}{\sfdefault}{m}{sl}
\SetMathAlphabet{\mathsfit}{bold}{\encodingdefault}{\sfdefault}{bx}{n}